\documentclass[twocolumn]{aastex631}
\usepackage{multirow}
\usepackage{cancel}

\begin{document}

\title{Warm Ionized Gas Outflows in Active Galactic Nuclei: What Causes it?}

\correspondingauthor{Payel Nandi}
\email{payel.nandi@iiap.res.in}

\author[0000-0002-0786-7307]{Payel Nandi}
\affiliation{Indian Institute of Astrophysics, Block II, Koramangala, Bangalore, 560034, India}
\affiliation{Joint Astronomy Programme, Department of Physics, Indian Institute of Science, Bangalore, 560012, India}

\author[0000-0002-4998-1861]{C.S. Stalin}
\affiliation{Indian Institute of Astrophysics, Block II, Koramangala, Bangalore, 560034, India}

\author[0000-0002-4464-8023]{D.J.Saikia}
\affiliation{Inter-University Centre for Astronomy and Astrophysics, IUCAA, Pune 411007, India}



\begin{abstract}
The driving force behind outflows, often invoked to understand the correlation between the supermassive black holes powering active galactic nuclei (AGN) and their host galaxy properties, remains uncertain. We provide new insights into the mechanisms that trigger warm ionized outflows in AGN, based on findings from the MaNGA survey. Our sample comprises 538 AGN with strong [OIII]$\lambda$5007 emission lines, of which 197 are detected in radio and 341 are radio-undetected. We analyzed [OIII]$\lambda$5007 line in summed spectra, extracted over their central 500$\times$500 pc$^2$ region. The calculated  Balmer 4000 $\AA$ break, D$_n$4000 is larger than 1.45 for $\sim$95$\%$ of the sources indicating that the specific star-formation rate in their central regions is smaller than 10$^{-11.5}$ yr$^{-1}$, pointing to evidence of negative AGN feedback suppressing star-formation. Considering the whole sample, radio-detected sources show an increased outflow detection rate (56$\pm$7\%) compared to radio-undetected sources (25$\pm$3\%). They also show higher velocity, mass outflow rate, outflow power and outflow momentum rate. We noticed a strong correlation between outflow characteristics and bolometric luminosity in both samples, except that the correlation is steeper for the radio-detected sample. Our findings suggest (a) warm ionized outflows are prevalent in all types of AGN, (b) radiation from AGN is the primary driver of these outflows, (c) radio jets are likely to play a secondary role in enhancing
the gas kinematics over and above that caused by radiation, and (d) very low star-formation in the central regions of the galaxies, possibly due to negative feedback of AGN activity.
\end{abstract}


\section{Introduction} \label{sec:intro}
Supermassive black holes (SMBHs) with masses (M$_{BH}$) greater than 10$^6$ M$_{\odot}$ are generally known to reside at the centres of all massive galaxies in the Universe (\citealt{2018Natur.553..307M}; and references therein). 
A small fraction of these galaxies hosts active galactic nuclei (AGN) caused by the accretion of matter from the surroundings by these SMBHs \citep{2017A&ARv..25....2P}, though less common in dwarf galaxies that host intermediate-mass black holes with M$_{BH}$ $<$ 10$^6$ M$_{\odot}$ \citep{2022NatAs...6...26R}. The process of accretion which leads to the release of enormous amounts of energy in the form of radiation \citep{1969Natur.223..690L} as well as particles via relativistic jets \citep{2009MNRAS.395..518C} is believed to affect their host galaxies via a process called feedback. AGN feedback is invoked to explain the observed correlation between
M$_{BH}$ and various host galaxy properties (\citealt{2023NatAs...7.1376Z} and references therein).

A viable feedback mechanism in AGN is outflows. 
These outflows are dynamic phenomena, representing 
the expulsion of vast amounts of matter and energy from the vicinity of 
SMBHs at the centres of galaxies. 
They play a crucial role in shaping the 
surrounding environment and influencing the evolution of galaxies 
\citep{2023Natur.624...53G}. They  are multifaceted, as seen in 
molecular, neutral and ionized gas \citep{2023ApJ...959..116N, 2023MNRAS.521.1832R, 2023MNRAS.520.5712S, 2023Sci...382..554I} and can occur on 
various scales, spanning from relatively small-scale winds to colossal jets 
extending over intergalactic distances \citep[][for a review]{2014MNRAS.441.3306H,2023ApJ...959..116N,2023Sci...382..554I,
2023Natur.624...53G,2022JApA...43...97S}. 
They can profoundly impact the galaxy's evolution by regulating the
rate of star formation, distributing elements crucial for planetary systems, 
and even influencing the growth of SMBHs  \citep{ 2023ApJ...950...81N, 2023A&A...678A.127V, 2024ApJ...973....7N}.
Irrespective of outflows being prevalent in AGN, questions such as
(a) what drives these outflows and (b) at what scales they operate 
are not conclusively known and are highly debated. 
The potential mechanisms that could drive these outflows could be
radiation and/or radio jets \citep{2018NatAs...2..181W}. 

Studies are available in the literature aimed at identifying the main driving 
mechanisms for outflows. They are focused on individual systems as well as on a sample
of sources. For example, the correlation noticed between the ionized [OIII]$\lambda$5007 gas outflow and the radio jet in NGC 1068, seems to favour a jet driven outflow \citep{2014A&A...567A.125G,2021A&A...648A..17V}. 
In IC 5063, \cite{2015A&A...580A...1M} found evidence of molecular, atomic
and ionized outflows and conclude that both the radiation and jet could drive the outflow, however, the jet being the dominant driver. Other studies that find favour of
jets triggering outflows  include that of NGC 1337 \citep{2016A&A...590A..73A}, 
a sample of 10 quasars at $z$ $<$ 0.2 \citep{2019MNRAS.485.2710J}, 
3C 273 \citep{2019ApJ...879...75H}, 
ESO 420$-$G13 \citep{2020A&A...633A.127F}, NGC 5643 \citep{2021A&A...648A..17V}, 
NGC 1386 \citep{2021A&A...648A..17V},
J1316+1753 \citep{2022MNRAS.512.1608G}, B2 0258+35 \citep{2022NatAs...6..488M}, 
N0945+1737 \citep{2022A&A...665A..55S}, the Teacup galaxy \citep{2023A&A...671L..12A}
and the dwarf AGN NGC 4395 \citep{2023ApJ...959..116N}. 
However, observations of Mrk 231 support radiation
from accretion driving the outflow \citep{2015A&A...583A..99F}.
While the above studies are focused on using high-resolution observations, 
low spatial resolution studies do exist.
\cite{2013MNRAS.433..622M} from analysis of a larger sample of quasars 
using the Sloan Digital Sky Survey (SDSS)\footnote{https://sdss.org/} spectra, found that sources  more luminous in
radio band tend to have broader [OIII]$\lambda$5007 line profile, while,   
\cite{2014MNRAS.442..784Z} found favour of outflows being driven 
via radiative output from quasars. 
From an analysis of SDSS spectra of Type 2 AGN,
\cite{2016ApJ...817..108W} found that while outflows are prevalent in
Type 2 AGN, they are not directly related to radio activity.
Alternatively, from an analysis of the SDSS spectra
of radio AGN, \cite{2023A&A...674A.198K} found that radio
jets are more effective in driving outflows when they are young.
\cite {2019A&A...631A.132M} too found that the chance of finding outflows is more in 
compact radio sources possibly hosting young radio jets.
Recently, from an
analysis of a large sample of AGN, \cite{2024MNRAS.528.3696L}
found the outflow velocity to correlate with radio power. 
However, \cite{2023ApJ...954...27A}, conclude, that both 
accretion  and  radio activity can have a role in
driving outflows. 

These studies aimed at finding the driver of outflows, have focussed on both radio-loud and radio-quiet quasars.  This is misleading as even radio-quiet quasars are found to have radio jets that could
impact the outflow. Also, the separation of quasars into radio-loud and radio-quiet 
is questioned, and an alternative  division of AGN into jetted and non-jetted 
sources is proposed \citep{2017NatAs...1E.194P}. Thus, the driver for outflows in AGN remains unsettled. To overcome 
the above limitations, firstly, we utilised in this work a sample of AGN separated  
into radio-detected and radio-undetected. This inclusion of radio-undetected sources 
makes sure that these sources lack clear signatures of radio jets (considering that the 
radio emission in radio-detected AGN is due to jets in them and not due to star formation in their hosts) at the sensitivity levels of existing radio surveys. And, secondly, we used 
spatially resolved spectroscopic data on a large sample of AGN hitherto not utilised 
for such a comparative study.

\section{Sample} \label{sec:samp}
Our initial sample of sources was derived from the MaNGA (Mapping Nearby Galaxies at Apache Point Observatory; \citealt{2015ApJ...798....7B}) survey, a spectroscopic program under SDSS-IV. MaNGA employs a fiber-based integral field unit (IFU) spectroscopic technique, utilizing the two BOSS spectrographs mounted on the 2.5-meter Sloan Foundation Telescope at Apache Point Observatory.
MaNGA has observed 10,010 unique galaxies with redshifts ranging from 0.01 to 0.15 \citep{2017AJ....154...86W}, using different IFU configurations. The spatial resolution achieved is between \(2^{\prime\prime}\) and \(2.5^{\prime\prime}\), with a spectral resolution of approximately 2000. The program’s field of view varies from \(12^{\prime\prime}\) to \(32^{\prime\prime}\), depending on the IFU configuration, covering a spatial range of 1.5 to 2.5 effective radii of the observed galaxies. We cross-correlated the sources in the MaNGA catalogue \citep{2017AJ....154...86W}
with the latest version of the Million Quasars Catalogue 
(MILLIQUAS; \citealt{2023arXiv230801505F}) to identify genuine AGN in the 
MaNGA catalogue, using a search radius of 2$^{\prime\prime}$. MILLIQUAS is
a collection of all published AGN and quasars till 30 June 2023, amounting
to a total of 1,021,800 sources. Our cross-correlation of MaNGA sources with
MILLIQUAS led to a sample of  1,142 AGN. As these sources were pulled from various
surveys in MILLIQUAS, we checked the position of these 1142 sources in the Baldwin-Phillips-Terlevich diagram (BPT; \citealt{1981PASP...93....5B}) for homogeneity.  We took an aperture of 500$\times$500 square pc box centred on the source and calculated the flux values of  [OIII]$\lambda$5007, H$\beta$, H$\alpha$, [NII]$\lambda$6584, [SII]$\lambda$6718 and [SII]$\lambda$6732 lines from the Data Analysis Pipeline (DAP) products \citep{2019AJ....158..231W} of MaNGA. Then, we plotted the flux ratio between  [OIII]$\lambda$5007 and H$\beta$, [SII]$\lambda$(6717+6732) and H$\alpha$  and also the flux ratio between  [OIII]$\lambda$5007 and H$\beta$, [NII]$\lambda$6584 and H$\alpha$ in for the sources in the BPT diagrams. Then, out of these 1142 sources, 718 are above the star-forming line in the [SII]$\lambda$(6717+6732)/H$\alpha$ vs [OIII]$\lambda$5007/H$\beta$ diagram and 740 sources lie above the star formation line in the [NII]$\lambda$6584/H$\alpha$ vs [OIII]$\lambda$5007/H$\beta$ diagram with 688 common sources in both the BPT diagrams. We considered these 688 AGN for our analysis out of which 252 are Seyferts, and 436 are LINERs.

We cross-matched these 688 AGN with the VLA Faint Images of the Radio Sky at Twenty-Centimeters (FIRST) survey \citep{1995ApJ...450..559B} using an angular separation of 3$^{\prime\prime}$. The FIRST survey, conducted with the NRAO Very Large Array in its B-configuration, provides radio maps of the sky at 20 cm (1.4 GHz) with a beam size of approximately 5.4$^{\prime\prime}$, a typical root mean square (rms) noise level of 0.14 mJy beam$^{-1}$. Through this cross-matching process, we identified 217 AGN with radio counterparts in the FIRST catalogue, exhibiting flux densities greater than 0.5 mJy, classifying them as radio-detected. The remaining 471 AGN, lacking radio counterparts in the FIRST survey, were categorized as radio-undetected.
Out of these 471 sources, 18 sources are not 
covered by the FIRST survey. Not considering those 18 sources, our final 
radio-undetected sample consists of 453 sources. Of the 217 radio-detected sample 
95 sources are Seyfert type AGN while 122 are LINERs. Similarly, among the  453 radio-undetected sample,  149 sources are Seyfert type AGN while 304 sources 
are LINERs. The positions of these sources in the BPT diagrams are shown in Appendix \ref{app-bpt-samp-line} (see Fig. \ref{fig:bpt}).  Both the radio-detected and the radio-undetected samples 
have similar distributions in the redshift and optical B-band brightness plane (see Fig. \ref{fig:lum-z}).
A Kolmogorov-Smirnov (KS) test carried out on their distributions of redshift and B-band brightness
indicates that the two samples are indeed indistinguishable with statistics of 
0.07 and a p-value of 0.12.

\section{Analysis} \label{sec:annal}
We focussed our analysis on a total of 217 AGN with radio detection and 453 AGN 
without radio detection in the FIRST survey. For this, we used the data reduction pipeline (DRP) products 
\citep{2016AJ....152...83L} LOGCUBE of SDSS DR17. The DRP products contain the processed, and calibrated spectra for each spaxel in the field of view in form of cube for each source.

For each of the sources studied in this work, we generated summed spectra in 
the rest frame of the sources over a square with the length of the side of 500 pc. The choice of 500 pc is due to our requirement of having at least one spaxel to generate
the spectra for most of the sources.

We fitted the [OIII]$\lambda$5007 profile with multiple Gaussian components along with a first-order polynomial for the continuum using the nonlinear least 
square fitting algorithm within the {\it curvefit} module in the {\it Scipy} library. We fitted [OIII]$\lambda$5007 because we are only interested in the warm ionised phase of outflow that is traced by the forbidden bright line [OIII]$\lambda$5007. During the fit, we kept the width, peak and amplitude 
of each of the components as free parameters. Also, we restricted the fitting to 
those sources for which the signal-to-noise ratio (SNR) of the line is more than 3.0. 
Here, SNR refers to the ratio of the flux at the peak of the line to the standard 
deviation of continuum fluxes on either side of the line. We used a total of 
80$~$\AA ~width for the spectral region (4977$-5057$~\AA)  during the fitting. For some sources, the [OIII]$\lambda$5007 line profile was adequately modelled with a single Gaussian component. However, in cases where the residual, defined as (data - model)/data, exceeded 10\%, additional Gaussian components were considered. The residual was reassessed after each addition and compared to the previous fit. If including an additional Gaussian reduced the residual and smoothed fluctuations in both the surrounding continuum and the line region, the extra component was adopted. Otherwise, the fit was restricted to the minimum number of Gaussian components required. For cases where more than one Gaussian component was necessary, an additional criterion was applied: the peak of the second and third components have to exceed three times the standard deviation of the continuum fluxes to confirm their statistical significance. In the radio-detected sample, the SNR of the first outflow
component ranges from 10 to 306, while the SNR of the second outflow component ranges
from 8 to 331. In the radio-undetected sample, the SNR of the first outflow component ranges from 4 to 199, and the SNR of the second outflow component ranges from 3 to 122.
We show in Fig. \ref{fig:fitting}, the spectral fits to three sources, one requiring a single Gaussian component, while the others requiring two and three Gaussian components, respectively. We also manually inspected each of the fitted spectra to ensure their fitting was correct. 

After the fitting, we corrected the measured outflow fluxes 
for galactic extinction using \cite{1989ApJ...345..245C} and the E(B$-$V) values given in the header of DRP files. The fluxes were also corrected for internal extinction using the  H$\alpha$ and H$\beta$ ratio taken from DAP products and following \cite{1972ApJ...172..593M, 1995ApJS...98..171V} and \cite{2000ApJ...533..682C}. Under the theoretical assumption of case B recombination, the intrinsic H\(\alpha\)/H\(\beta\) ratio was taken as 3.1 \citep{2006agna.book.....O}.

\section{Results and Discussion} \label{sec:result}

\subsection{Detection of outflows}\label{sec:outflow-detec}
For both the radio-detected and the radio-undetected samples, we searched for 
the signature of outflows over a region of  500 $\times$ 500 square pc
centered on each of the sources. The [OIII]$\lambda$5007 line was detected at 
the 3$\sigma$ limit for 197 sources in the radio-detected category and 341 
sources in the radio-undetected category. Our final sample thus consists of 
a total of 538 sources with strong [OIII]$\lambda$5007 for further analysis. 

For the radio-detected sample, out of 197 sources, for 86 sources, a single Gaussian component proved sufficient for fitting the [OIII]$\lambda$5007 line, for 77 sources, two Gaussian components were needed to fit the line, 
while for 34 sources three Gaussian components were needed to model the line.
Considering sources that require more than one Gaussian component to 
well represent the [OIII]$\lambda$5007 line,  in the radio-detected category, a 
total of 111 out of 197 sources (56$\pm$7\%), prominently showed discernible 
signatures indicative of outflows. 

Similarly, in the  radio-undetected sample, out of the 341 sources, 257 sources 
required a single Gaussian component to well represent the [OIII]$\lambda$5007 line,  
while 69 sources required two Gaussian components and 15 sources needed three 
Gaussian components to well model the line. Thus, in  the radio-undetected 
sample,  we detected outflows for 84 sources i.e. 25$\pm$3\% of the sources showed signatures of outflows. This clearly indicates that the outflow detection rate is higher in the radio-detected sample compared to the radio-undetected sample.

\begin{figure}
    \centering
    \includegraphics[scale=0.40]{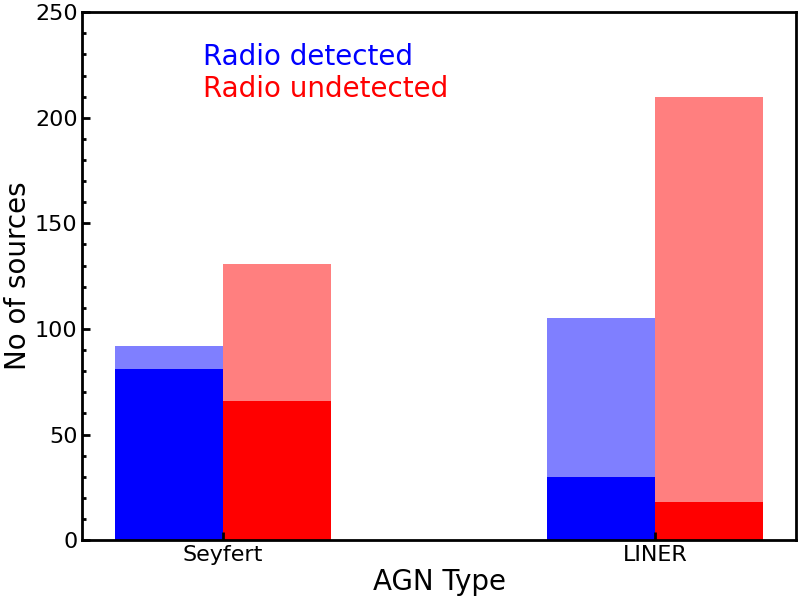}
    \caption{Bar chart of the different types of AGN (Seyferts and LINERs) used in this study. Here, the large blue bar refers to the total radio-detected sources, and the large red bar refers to the total radio-undetected sources. Dark-shaded regions refer to outflow-detected sources.}
    \label{fig:agn_type}
\end{figure}

We also classified our sample into Seyferts and LINERs based on their location in the BPT diagram (Fig. \ref{fig:bpt}) to investigate their prevalence of outflows.  In the radio-detected sample, we have 92 Seyferts and 105 LINERs for which [OIII]$\lambda$5007 was significantly detected. Of these,  we detected outflows in 81 Seyferts and 30 LINERs. Similarly, in the radio-undetected sample, we detected [OIII]$\lambda$5007 line in 131 Seyferts and 210 LINERs. Of these, we detected outflows in 66 Seyferts and 18 LINERs. The results of the analysis are summarised in Table \ref{tab:det} and in Fig. \ref{fig:agn_type}. 

Considering the total sample, we found that 66$\pm$7\% of Seyferts show outflows, compared to only 15$\pm$2\% LINERs. In radio-detected Seyferts, outflows are detected in 88$\pm$13\% sources, while in radio-detected LINERs, it is 29$\pm$6\%. In the case of the radio-undetected sample, we detected outflows for 50$\pm$8\% Seyferts and 8$\pm$2\% LINERs. Thus Seyferts consistently show more frequent outflows than LINERs, both in radio-detected and radio-undetected samples. This is in agreement with a recent study by \cite{2024AJ....168...37T} who, too, from a systematic analysis of the SDSS spectra of a large sample of Seyferts and LINERs, found that the probability of detecting outflows in LINERs is lower compared to that of Seyferts. 

\begin{table*}
    \centering
    \caption{Summary of the sources analysed for ionised outflows in the  [OIII]$\lambda$5007 line.}
    \begin{tabular}{lcccccccc}
    \hline
     & \multicolumn{2}{c}{Total Sample} & \multicolumn{3}{c}{radio-detected} & \multicolumn{3}{c}{radio-undetected} \\
     &  Seyferts  & LINERs & Total & Seyferts & LINERs  & Total  & Seyferts  & LINERs \\
        \hline 
Total number of sources                             & 223  & 315   & 197 & 92 & 105 & 341 & 131 & 210 \\
Number of source with one Gaussian component        & 76   & 267   &  86 & 11 &  75 & 257 &  65 & 192 \\
Number of sources with two Gaussian components      & 99   & 47    &  77 & 47 &  30 &  69 &  52 &  17 \\
Number of sources with three Gaussian components    & 48   &  1    &  34 & 34 &   0 &  15 &  14 &   1 \\
Number of sources with outflow                      & 147  & 48    & 111 & 81 &  30 &  84 &  66 &  18 \\
Outflow detection rate in percentage                & 66$\pm$7   & 15$\pm$2    &  56$\pm$7 & 88$\pm$13 &  29$\pm$6 &  25$\pm$3 &  50$\pm$8 &  8$\pm$2 \\
    \hline 
    \end{tabular}
    \label{tab:det}
\end{table*}

\subsection{Comparison of kinematics properties of outflow} \label{sec:outflow_kine}
In the following sections, we compared the kinematic properties of outflow in our sample of sources. In cases where two outflowing components were detected, 
we initially consider only the brightest of the two in the analysis of the
kinematic properties of the outflow. We later also considered the outflowing component of lower brightness, and also the one with the higher velocity. The values for the brighter component are tabulated in 
Table \ref{tab:out_high_lum} for the total sample, the Seyferts and
LINERs and among them separately into radio-detected and radio-undetected samples. 
Similar Tables for the other two cases are given in the Appendix B.

\subsubsection{Velocity shift}\label{sec:Vshift}
We measured the velocity shift (V$_{shift}$) of the outflowing component relative 
to the narrow component of the [OIII]$\lambda5007$ emission line.
In this definition, a negative value of $V_{shift}$ corresponds to the broad blueshifted component and a positive value of $V_{shift}$ corresponds to the broad redshifted component. The distribution of V$_{shift}$  for both the radio-detected and radio-undetected sample of sources is shown in the upper left panel of Fig. \ref{fig:kine}. The  KS test reveals that the distributions of the two samples are not statistically different, with a p-value of 0.4. For the radio-detected sample,  V$_{shift}$ ranges from $-$782 km s$^{-1}$ to 463 km s$^{-1}$, with a mean of $-$178 km s$^{-1}$ and a median of $-$163 km s$^{-1}$ with median uncertainty of 46 km s$^{-1}$. Similarly for the radio-undetected sample, V$_{shift}$ ranges from $-$628 km s$^{-1}$ to 108 km s$^{-1}$, with a mean of $-$234 km s$^{-1}$ and a median of $-$167 km s$^{-1}$ with median uncertainty of 70 km s$^{-1}$.
In both our samples, we found more sources ($\sim$ 80\%) to show blue asymmetries of their [OIII]$\lambda$5007 line relative to red asymmetry. This could be because of the redshifted part of the bipolar outflow being obscured by dust and/or seen at a lower S/N than the blue-shifted component and thus undetected \citep{2024Natur.630...54B}.

\subsubsection{Velocity dispersion}\label{sec:FWHM_out}
We parameterise the emission line profile of the outflowing component using
the dispersion of the line parameter ($\sigma_{measured}$) obtained from the fitting of the [OIII]$\lambda5007$ line. This measured velocity dispersion of the outflowing component,  $\sigma_{measured}^2$ = $\sigma_{out}^2$ + $\sigma_{inst}^2$, 
where $\sigma_{out}$ and  $\sigma_{inst}$ are the intrinsic velocity dispersion of
the outflowing component and the dispersion of the instrumental line
spread function. For MaNGA, the 1$\sigma$ width of the instrumental line 
spread funciton is $\sim$70 km s$^{-1}$ \citep{2022ApJ...928...58L}. 
To estimate $\sigma_{out}$, we subtracted $\sigma_{inst}$ from $\sigma_{measured}$
in quadrature.  From the estimated $\sigma_{out}$, we calculated
the full width at half maximum (FWHM) of the outflowing component as
$FWHM_{out} = 2\sqrt{2 ln 2}$ $\sigma_{out} = 2.35\sigma_{out}$, which is 
valid for a true Gaussian profile. 
The distribution of $FWHM_{out}$ is shown in the upper middle panel of 
Fig. \ref{fig:kine} for both the radio-detected and radio-undetected samples.

For the radio-detected category, FWHM$_{out}$ ranges from 169 km s$^{-1}$ to 1398 km s$^{-1}$, with a mean value of 646 km s$^{-1}$ and a median of 626 km s$^{-1}$ with median uncertainty of 59 km s$^{-1}$. 
For the radio-undetected category, FWHM$_{out}$ ranges from 140 km s$^{-1}$ to 1171 km s$^{-1}$ and it is lower compared to radio-detected sample, with a mean of 526 km s$^{-1}$ and a median of 518 km s$^{-1}$ along with median uncertainty of 82 km s$^{-1}$. Therefore, this larger range and higher values of FWHM$_{out}$ for radio-detected sources indicate that the outflowing material in them is more kinematically disturbed compared to radio-undetected sources.

\subsubsection{Outflow velocity}\label{sec:Vout}
We define the outflow velocity, \( V_{out} \), as the sum of the velocity 
difference ($|V_{shift}|$) between the outflowing component and the narrow 
component, plus two times the standard deviation of the outflowing component 
($\sigma_{out}$), i.e., \( V_{out} = |V_{shift}| + 2\sigma_{out} \) 
\citep{2024arXiv240719008P}. The distribution of \( V_{out} \) is shown in the upper right panel of Fig. \ref{fig:kine} for both the radio-detected and radio-undetected samples. The KS test indicates that the distributions of the radio-detected and radio-undetected populations are statistically distinct, with a p-value of 6$\times$10$^{-3}$. This low p-value suggests that the likelihood of these two distributions being drawn from the same parent population is very low. 

For the radio-detected sample, the outflow velocity \( V_{out} \) spans from 271 km s\(^{-1}\) to 1970 km s\(^{-1}\), with an average velocity of 788 km s\(^{-1}\) and a median velocity of 705 km s\(^{-1}\) with median uncertainty of 74 km s$^{-1}$. In contrast, the radio-undetected sources exhibit a range of \( V_{out} \) from 118 km s\(^{-1}\) to 1387 km s\(^{-1}\), with a lower mean velocity of 691 km s\(^{-1}\) and a median of 610 km s\(^{-1}\) along with median uncertainty of 98 km s$^{-1}$. The higher velocities in the radio-detected sample may imply that radio emission is linked to more powerful or sustained outflows, possibly associated with jet-driven mechanisms or enhanced AGN activity. The contrast in median and mean velocities between the two samples supports the idea of a significant difference in outflow dynamics related to the presence of radio emission.

\subsubsection{Asymmetric index}\label{sec:AI}
To evaluate the asymmetry of the total [OIII]$\lambda$5007 line profile, we 
utilize the asymmetry index (AI). Following \cite{2014MNRAS.442..784Z}, the 
AI is defined as 
\begin{equation}
    AI = \frac{(V95 - V50) - (V50 - V05)}{V95 - V05}
\end{equation}
Here V95, V50 and V05 are the velocities at which 95\%, 50\% and 5\% of the 
emission line flux is found.  A value of zero indicates a symmetric profile, a 
positive value suggests redshifted wings, and a negative value indicates 
blueshifted wings. The middle left panel of Fig. \ref{fig:kine} displays the distribution of AI values for both the radio-detected and radio-undetected samples. Statistically, these distributions differ, with a KS test statistic of 0.32 and a p-value of 0.04, suggesting a significant but moderate distinction between the two groups.

For the radio-detected sample, AI values range from $-$0.46 to 0.12, with an average of $-$0.15 and a median of $-$0.16 with median uncertainty of 0.05. In contrast, the radio-undetected sample has a wider range from $-$0.51 to 0.16, with a mean value of $-$0.18 and a median of $-$0.17 with median uncertainty of 0.10. The consistently negative AI in both the samples suggests the dominance of the blueshifted component of the bipolar outflow relative to the redshifted component of the outflow, as discussed in section \ref{sec:Vshift}.

\subsubsection{Outflow mass}
We determined the mass of the outflowing gas (M$_{out}$) following \cite{2024A&A...685A..99C} as 
\begin{equation}
M_{out} = 0.8 \times 10^8 \left(\frac{L_{[OIII]out}}{10^{44} erg s^{-1}}\right) \left(\frac{500 cm^{-3}}{n_e}\right) \left(\frac{Z_{\odot}}{Z}\right) M_{\odot}
\end{equation}
Here, $L_{[OIII]out}$ is the luminosity of the outflowing component calculated 
from the flux of the outflowing component of [OIII]$\lambda5007$ and corrected for dust extinction following the procedure given in Section 3. The outflow mass also depends on both the electron density (n$_e$) and the gas phase metallicity of the medium. We calculated n$_e$ using the ratio of [SII]$\lambda$6718 to [SII]$\lambda$6732, assuming an electron temperature of 10,000 K, using \textit{pyneb} \citep{2015A&A...573A..42L}. We also determined the gas phase metallicity using [OIII]$\lambda\lambda$4959,5007, 
[NII]$\lambda\lambda$6548,6584, and the Balmer lines (H$\alpha$ and H$\beta$), 
as described by \cite{2022MNRAS.513..807D}.

The distribution of M$_{out}$ for both the radio-detected and radio-undetected samples is illustrated in the middle panel of Fig. \ref{fig:kine}. A KS test shows that the two distributions are statistically distinct, with a p-value of 6$\times$10$^{-4}$.
For the radio-detected sample, M$_{out}$ ranges from 181 M$_{\odot}$ to 2.4$\times$10$^{6}$ M$_{\odot}$, with a mean of 1.1$\times$10$^5$ M$_{\odot}$ and a median of 3.5$\times$10$^4$ M$_{\odot}$ with median uncertainty of 1.4$\times$10$^3$ M$_{\odot}$. 
 In contrast, the radio-undetected sample, shows M$_{out}$ values ranging from 30 M$_{\odot}$ to 1.5$\times$10$^5$ M$_{\odot}$, with a mean of 2.5$\times$10$^4$ M$_{\odot}$ and a median of 9.6$\times$10$^3$ M$_{\odot}$ along with median uncertainty of 8.6$\times$10$^2$ M$_{\odot}$. This finding suggests that outflow masses are notably higher in radio-detected sources than in radio-undetected sources.

\subsubsection{Mass outflow rate}
We calculated the mass outflow rate (\(\dot{M}_{out}\)), which represents the 
mass of gas outflowing per unit time as
\begin{equation}
\dot{M}_{out} = \frac{V_{out} M_{out}}{R}
\end{equation}
We considered R as  500 pc. The distribution of \(\dot{M}_{out}\) for both the radio-detected and radio-undetected samples is illustrated in the middle right panel of Fig. \ref{fig:kine}. According to the KS test, the distributions differ significantly, with a p-value of \(5 \times 10^{-5}\).
For the radio-detected sample, \(\dot{M}_{out}\) ranges from \(2.3 \times 10^{-4}\) to 5.1 M$_{\odot}$ yr$^{-1}$, with a mean of 0.17 M$_{\odot}$ yr$^{-1}$ and a median of 0.04 M$_{\odot}$ yr$^{-1}$ with median uncertainty of 0.004 M$_{\odot}$ yr$^{-1}$. For the radio-undetected sample, \(\dot{M}_{out}\) ranges from \(8.0 \times 10^{-5}\) to 0.2 M$_{\odot}$ yr$^{-1}$, with a mean of 0.03 M$_{\odot}$ yr$^{-1}$ and a median of 0.01 M$_{\odot}$ yr$^{-1}$ with median uncertainty of 0.002 M$_{\odot}$ yr$^{-1}$. These results indicate that \(\dot{M}_{out}\) is consistently higher in radio-detected sources compared to radio-undetected ones.

\subsubsection{Power of outflows}
The kinetic power of outflows (KP$_{out}$) is defined as 
\begin{equation}
     KP_{out} = \frac{1}{2} \dot{M}_{out} V_{out}^2 
\end{equation} 
The distribution of KP$_{out}$ for our sample of sources is shown in the bottom left panel of Fig. \ref{fig:kine}. The figure shows that radio-detected sources 
exhibit more powerful outflows than radio-undetected sources. From KS test, 
we found that the distributions of KP$_{out}$ for the radio-detected and radio-undetected samples are indeed different with a p-value of \(2 \times 10^{-7}\). 

For the radio-detected sample, KP$_{out}$ ranges from \(2.0 \times 10^{37}\) to \(1.8 \times 10^{42}\) erg s\(^{-1}\), with a mean of \(4.6 \times 10^{40}\) erg s\(^{-1}\) and a median of \(4.8 \times 10^{39}\) erg s\(^{-1}\) with median uncertainty of 1.8 $\times 10^{39}$ erg s$^{-1}$. In the radio-undetected sample, KP$_{out}$ varies from \(3.3 \times 10^{37}\) to \(7.0 \times 10^{40}\) erg s\(^{-1}\), with a mean of \(4.2 \times 10^{39}\) erg s\(^{-1}\) and a median of \(1.3 \times 10^{39}\) erg s\(^{-1}\) with median uncertainty of 5.9 $\times 10^{38}$ erg s$^{-1}$.

This indicates that radio-detected sources consistently exhibit higher outflow power compared to radio-undetected sources. This difference could suggest that radio emission is likely associated with more energetic outflows, possibly jets, potentially amplifying the AGN's feedback impact on the surrounding gas. Higher outflow power in radio-detected galaxies may be a sign of more efficient energy transfer from the AGN to the host galaxy's interstellar medium, possibly affecting star formation and the overall galactic environment \citep{2015ARA&A..53..115K}.

\subsubsection{Momentum rate of outflows}
\begin{figure*}
    \centering
    \vbox{
    \hbox{
    \hspace*{-0.4cm}\includegraphics[scale=0.32]{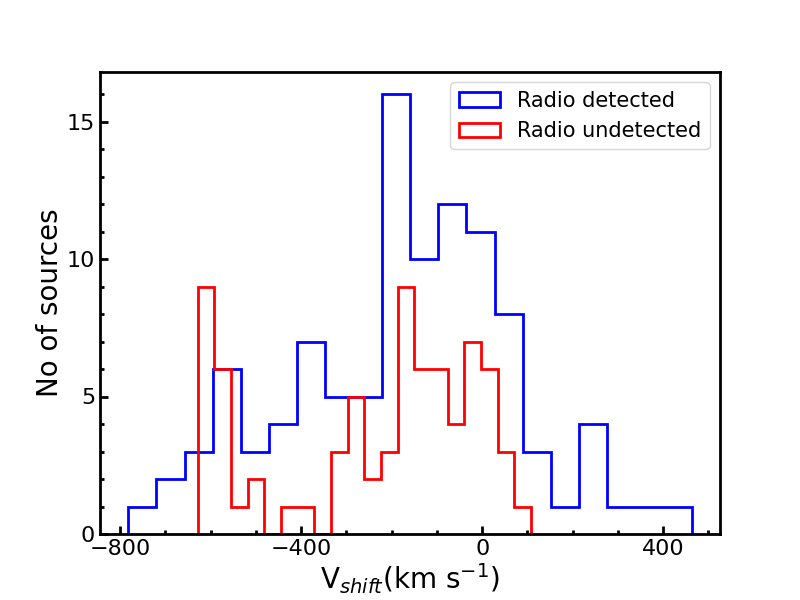}
    \hspace*{-0.6cm}\includegraphics[scale=0.32]{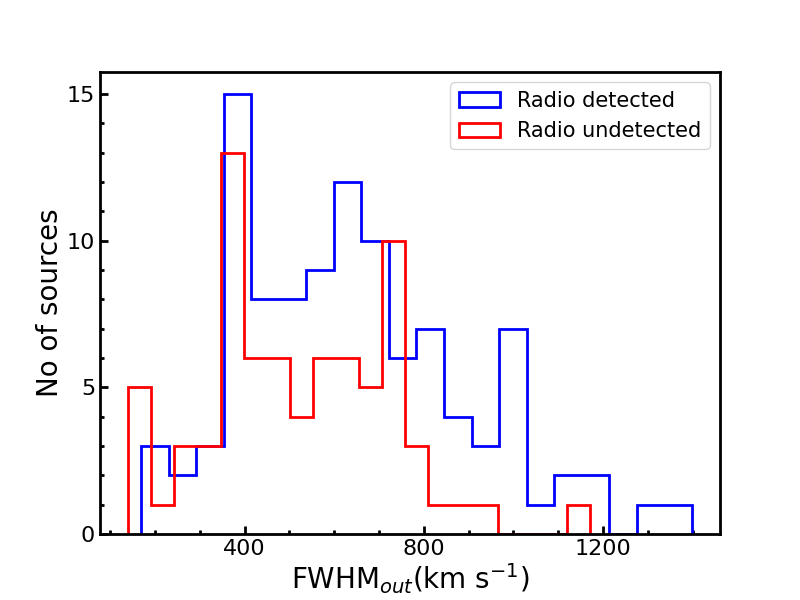}
    \hspace*{-0.6cm}\includegraphics[scale=0.32]{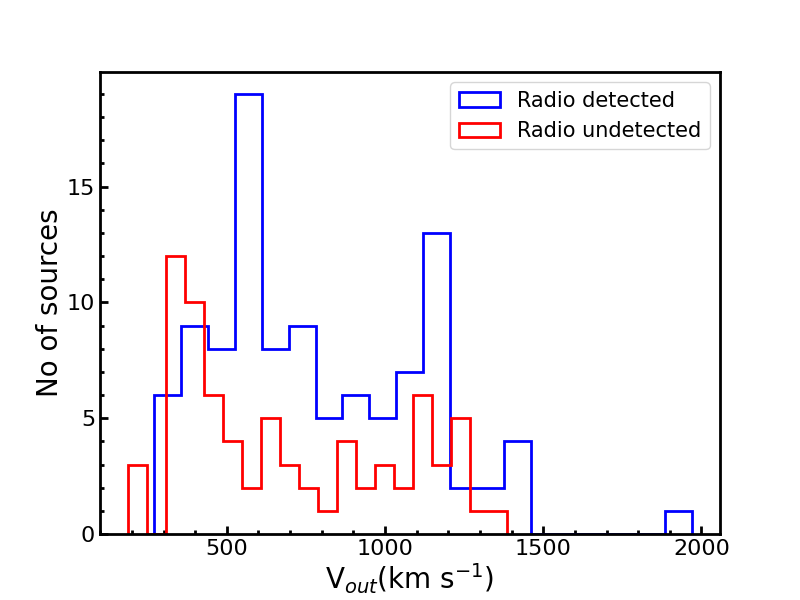}
    
    }
    \hbox{
    \hspace*{-0.4cm}\includegraphics[scale=0.32]{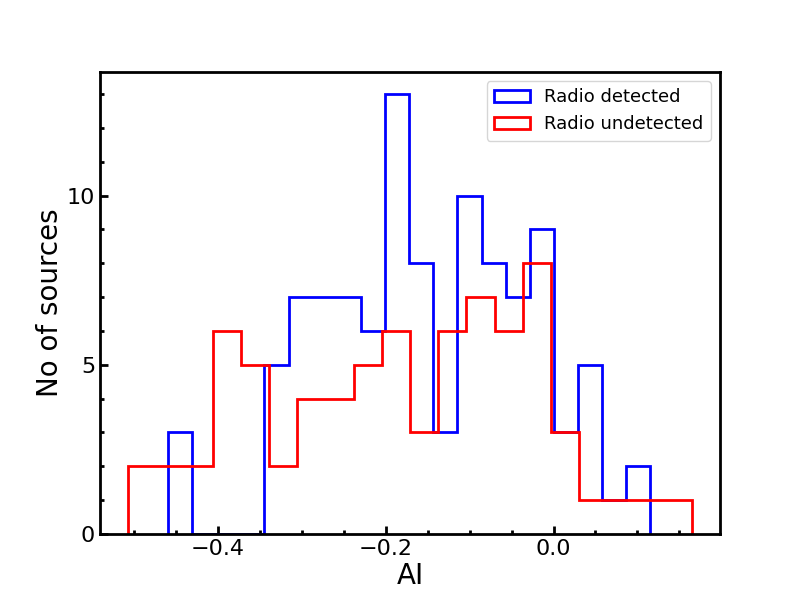}
    \hspace*{-0.6cm}\includegraphics[scale=0.32]{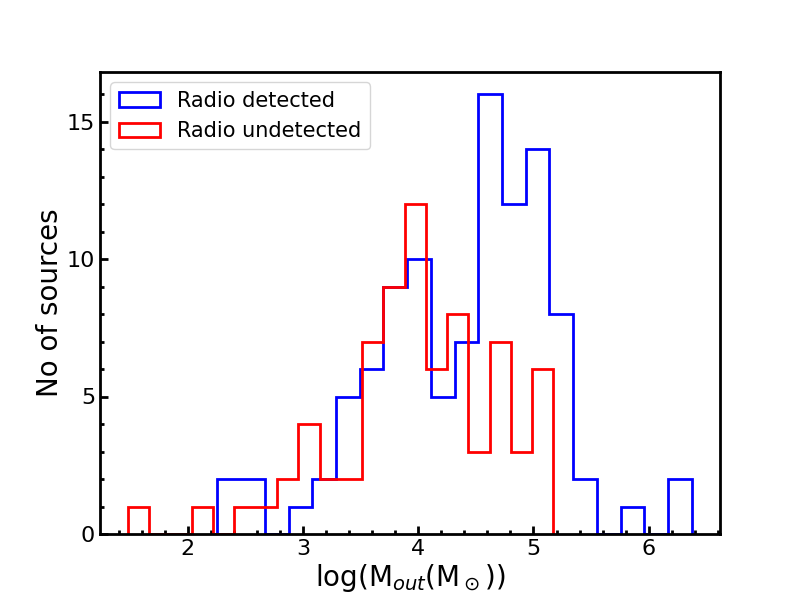}
    \hspace*{-0.6cm}\includegraphics[scale=0.32]{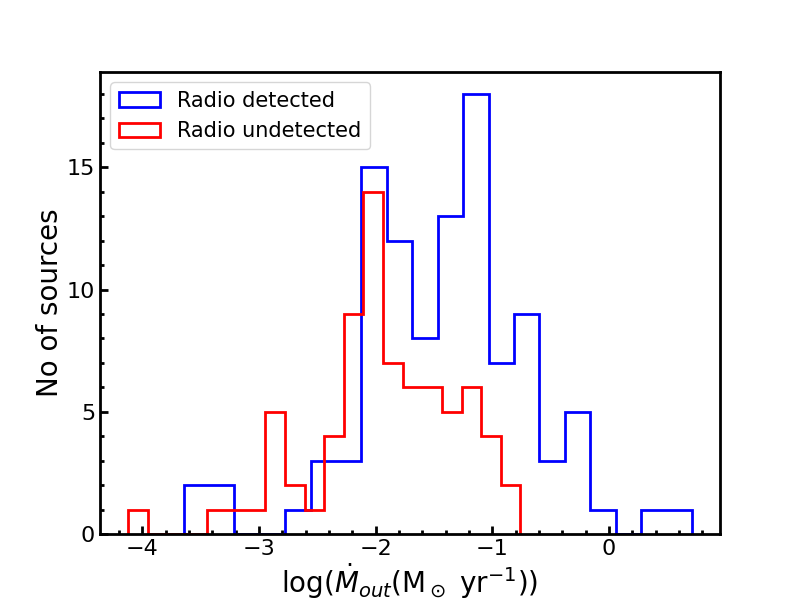}
    }
    \hbox{
    \hspace*{-0.4cm} \includegraphics[scale=0.32]{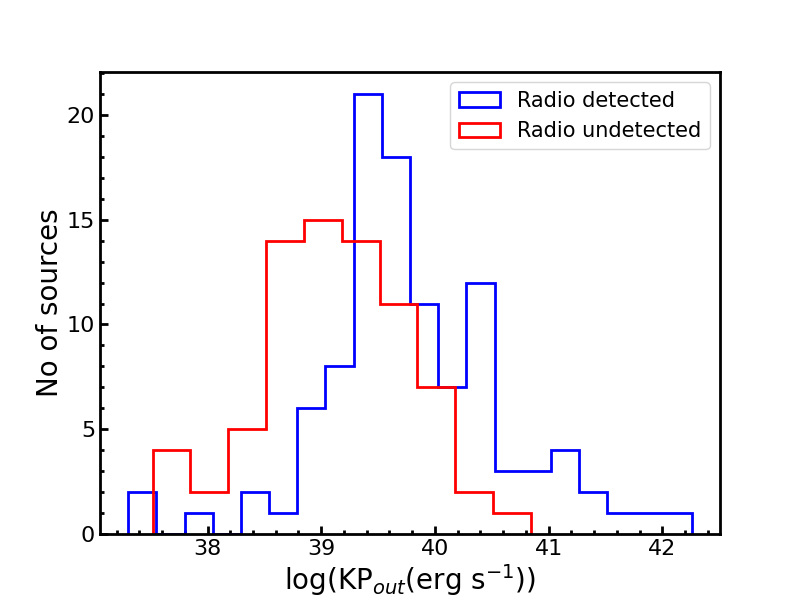}
    \hspace*{-0.6cm}\includegraphics[scale=0.32]{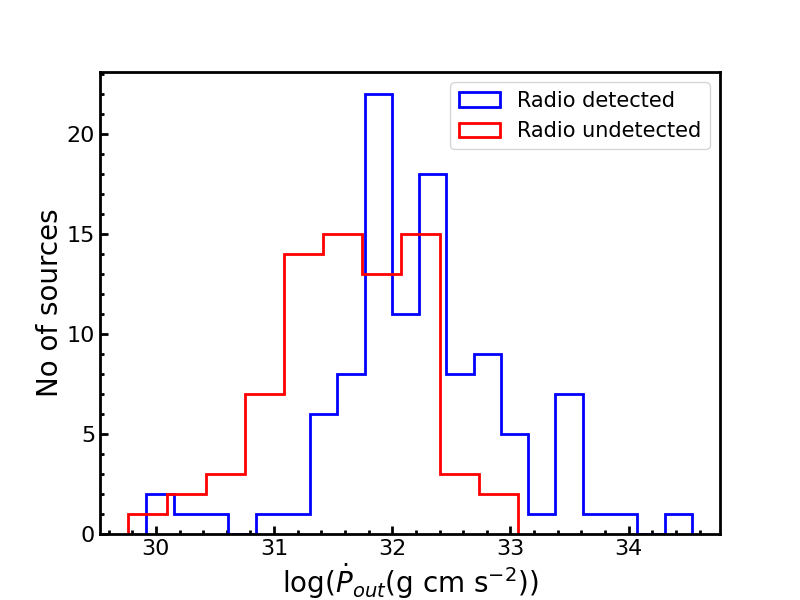}
    }
    }
    \caption{Histograms of different kinematics properties of outflows. The parameters are labelled in the respective plots. The blue and red histograms are for the radio-detected and radio-undetected samples, respectively.}
    \label{fig:kine}
\end{figure*}

The momentum rate of outflows (\(\dot{P}_{out}\)) is  defined as \(\dot{P}_{out} 
= \dot{M}_{out}V_{out}\). The distribution of \(\dot{P}_{out}\) (outflow momentum rate) for the sample is shown in the bottom middle panel of Fig. \ref{fig:kine}. According to the KS test, \(\dot{P}_{out}\) is significantly higher in radio-detected sources than in radio-undetected ones, with a p-value of \(9 \times 10^{-7}\). 

For radio-detected sources, \(\dot{P}_{out}\) spans from \(8.3 \times 10^{29}\) to \(3.4 \times 10^{34}\) g cm s\(^{-2}\), with a mean of \(9.2 \times 10^{32}\) g cm s\(^{-2}\) and a median of \(1.7 \times 10^{32}\) g cm s\(^{-2}\) with median uncertainty of 3.0$\times 10^{31}$ g cm s$^{-2}$. For the radio-undetected sources, \(\dot{P}_{out}\) ranges from \(5.8 \times 10^{29}\) to \(1.2 \times 10^{33}\) g cm s\(^{-2}\), with a mean of \(1.1 \times 10^{32}\) g cm s\(^{-2}\) and a median of \(5.1 \times 10^{31}\) g cm s\(^{-2}\) and with median uncertainty of 1.5$\times 10^{31}$ g cm s$^{-2}$. This indicates that \(\dot{P}_{out}\) is nearly an order of magnitude higher in radio-detected sources compared to radio-undetected sources.

The elevated \(\dot{P}_{out}\) in radio-detected sources suggests a stronger coupling between the AGN energy and the outflow momentum, enhancing feedback effects on the host galaxy. Given the higher outflow power (KP$_{out}$) and mass outflow rates (\(\dot{M}_{out}\)) in these sources, it appears that radio activity is correlated with more efficient AGN feedback. This combination of higher momentum, power, and mass flow may drive larger-scale gas movements, which could potentially lead to star formation suppression or triggering in the host galaxy more effectively than in radio-undetected sources \citep{2012ARA&A..50..455F, 2015ARA&A..53..115K, 2016A&ARv..24...10T}.

\begin{table*}
    \caption{Kinematic properties of the brightest outflows of Radio-detected sources (upper panel) and Radio-undetected sources (lower panel)}
    \label{tab:out_high_lum}
    \centering
    \label{tab:radio_out}
    \hspace*{-3cm}
    \begin{tabular}{lrrrrrrrrr}
    \hline
    Parameter & \multicolumn{3}{c}{Total} & \multicolumn{3}{c}{Seyferts} & \multicolumn{3}{c}{LINERs}\\
              & Range & Mean & Median & Range & Mean & Median & Range & Mean & Median \\
    \hline
     V$_{shift}$(km s$^{-1}$)  & $-$782 to 463 &  $-$178 &  $-$163 & $-$782 to 463 &  $-$106 &  $-$89 &  $-$695 to 258 &  $-$376 & $-$449 \\
     FWHM$_{out}$(km s$^{-1}$) &  169 to 1398 &  646 &  626 & 169 to 1398 & 635 & 611 & 257 to 1018 &  678 &  737 \\
     V$_{out}$(km s$^{-1}$)  & 271 to 1970 &  788 & 705 & 271 to 1970 &  720 & 642 & 378 to 1400 &  973 &  1084.0\\
     AI &  $-$0.46 to 0.12 &  $-$0.15 &  $-$0.16 & $-$0.34 to 0.12 &  $-$0.14 & $-$0.15 & $-$0.46 to 0.11 &  $-$0.18 &  $-$0.16 \\
     M$_{out}$(10$^2$ M$_{\odot}$) & 1.81 to 23681.85 & 1069.71 &  348.47 & 1.89 to 23681.85 &  1362.98 &  507.71 & 1.81 to 2120.98 & 273.70 & 41.16 \\
     $\dot{M}_{out}$(10$^{-3}$ M$_{\odot}$ yr$^{-1}$) & 0.23 to 5112.32 & 166.56 & 43.28 &  0.23 to 5112.32 & 212.13 & 59.45 & 0.33 to 424.39 & 42.87 & 8.02  \\
     KP$_{out}$(10$^{38}$ erg s$^{-1}$) &  0.20 to 18166.25 & 461.92 & 48.25 & 0.24 to 18166.25 & 591.12 & 55.63 & 0.20 to 1425.12 & 111.26 & 34.40 \\
     $\dot{P}_{out}$ (10$^{30}$ g cm s$^{-2}$) & 0.83 to 34222.09 & 923.43 & 165.39 & 0.83 to 34222.09 & 1177.20 & 198.87 & 0.91 to 2761.69 & 234.64 & 60.19 \\
    \hline
    \end{tabular}
    
    \label{tab:noradio_out}
     \hspace*{-3cm}
    \begin{tabular}{lrrrrrrrrr}
    \hline
    Parameter & \multicolumn{3}{c}{Total} & \multicolumn{3}{c}{Seyferts} & \multicolumn{3}{c}{LINERs}\\
              & Range & Mean & Median & Range & Mean & Median & Range & Mean & Median \\
    \hline
     V$_{shift}$(km s$^{-1}$)   & $-$628 to 108 & $-$234 & $-$167 & $-$628 to 108 & $-$173 & $-$134 & $-$624 to $-$211 & $-$526 & $-$585 \\
     FWHM$_{out}$(km s$^{-1}$)  & 140 to 1171 & 526 & 518 & 140 to 865 & 485 & 443 & 580 to 1171 & 720 & 672  \\
     V$_{out}$(km s$^{-1}$)     & 118 to 1387  & 691 & 610 & 188 to 1274 & 598 & 487 & 849 to 1387 & 1137 & 1136 \\
     AI   & $-$0.51 to 0.16 & $-$0.18 & $-$0.17 & $-$0.51 to 0.16 & $-$0.15 & $-$0.13 & $-$0.44 to $-$0.01 & $-$0.33 & $-$0.37   \\
     M$_{out}$(10$^2$ M$_{\odot}$) & 0.30 to 1488.24 & 252.60 & 96.10 & 4.70 to 1488.25 & 288.80 & 114.74 & 0.30 to 616.59 & 79.95 & 33.35 \\
     $\dot{M}_{out}$(10$^{-3}$ M$_{\odot}$ yr$^{-1}$) & 0.080 to 173.53 & 26.34 & 11.34 & 0.95 to 173.53 & 27.85 & 12.61 & 0.08 to 151.93 & 19.11 & 7.42 \\
     KP$_{out}$(10$^{38}$ erg s$^{-1}$) &  0.33 to 696.00 & 42.20 & 12.92 & 0.33 to 296.34 & 33.46 & 12.88 & 0.36 to 696.00 & 83.90 & 22.41 \\
     $\dot{P}_{out}$ (10$^{30}$ g cm s$^{-2}$) &  0.58 to 1154.74 & 106.05 & 50.95 & 1.99 to 793.52 & 98.54 & 49.45 & 0.58 to 1154.74 & 141.85 & 50.95 \\
         \hline
    \end{tabular}

\end{table*}

\

We have so far focused on the brightest outflowing component in this Section~\ref{sec:outflow_kine} when multiple components were detected. This approach may lead to an over representation of highly ionized outflows while potentially underestimating those with higher velocities but lower ionization. To assess this potential bias, we conducted two additional analyses: one considering the less luminous outflows and another focusing on the outflows with higher velocities. The results of these analyses are provided in Appendix \ref{app-table} (Tables \ref{tab:out_low_lum}, \ref{tab:out_high_speed}). Across all cases, we observed that radio-detected sources consistently exhibit higher velocities, mass outflow rates, outflow powers, and outflow momentum rates compared to their radio-undetected counterparts.

\subsection{Outflows in Seyferts vs LINERs}\label{out_sy_LINERS}
From Tables \ref{tab:out_high_lum}, \ref{tab:out_low_lum} and \ref{tab:out_high_speed}, it is evident that the FWHM$_{out}$ and V$_{out}$ are significantly greater in LINERs relative to Seyferts, suggesting higher outflow velocities in them. Additionally, a comparison of V$_{shift}$ reveals that the outflows are more blueshifted in LINERs than Seyferts. This larger velocity structure in LINERs may be linked to shock-dominated emission, as suggested by \cite{1995ApJ...455..468D}. However, when considering the mass outflow rate, outflow power and outflow momentum rate, Seyferts exhibit notably higher values, along with a greater outflow detection rate, as discussed in Section \ref{sec:outflow-detec}. This contrast is likely due to LINERs being at the low luminosity end of AGN, with ionizing power and accretion rate lower than that of Seyferts \citep{1980A&A....87..152H, 2008ARA&A..46..475H,2017FrASS...4...34M}, producing less outflowing material, thereby resulting in lower detection rates and less powerful outflows.

\subsection{Infrared properties of outflows}\label{sec:IR}

Of the sources analysed in this work, more than half of them are found to show 
outflows, as evidenced by the presence of shifted  broad asymmetric wings in their
[OIII]$\lambda$5007 line. Such observed line profile could be the result of
gas outflows from the central region of these sources \citep{2002ApJ...576L...9Z}. 
Such outflows can also be from the inner narrow line region (NLR) related to the winds from
AGN \citep{2000ApJ...545...63E}. Alternatively, outflows can also be driven by
star formation processes via winds from massive stars and/or Type 2 supernova
explosions \citep{2024arXiv240719008P}. Studies available in the literature point 
to infrared observations being an effective tool to distinguish
between these two processes, namely AGN driven and star formation driven outflows. Therefore, 
to investigate the infrared properties of the sources with outflows, we 
cross-correlated our sources with the  {\it Wide-field Infrared Survey Explorer} (WISE;\citealt{2010AJ....140.1868W}) catalog\footnote{\url{https://irsa.ipac.caltech.edu/Missions/wise.html}} 
using a search radius of 3 arcsec for both our samples. To ensure reliable 
data for analysis, we only included sources with a signal-to-noise ratio (SNR) 
greater than 3.0 in the W3 band. Since the WISE catalogue provides magnitudes in the 
Vega system by default, we converted them to the AB magnitude system following 
the guidelines provided at \url{https://wise2.ipac.caltech.edu/docs/release/allsky/expsup/sec4_4h.html}.

\begin{figure*}
    \centering
    \hbox{
    \hspace*{-0.5cm}\includegraphics[scale=0.33]{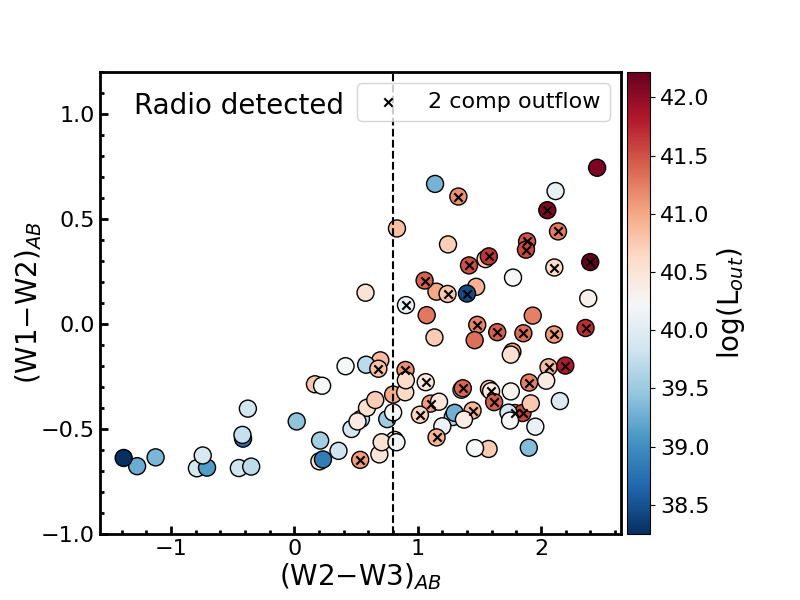}
    \hspace*{-0.5cm}\includegraphics[scale=0.33]{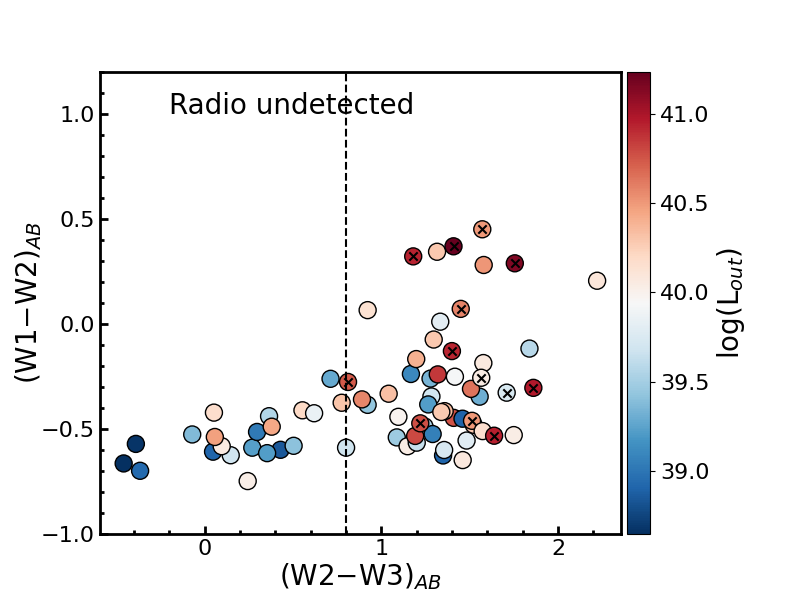}
    \hspace*{-0.6cm}\includegraphics[scale=0.33]{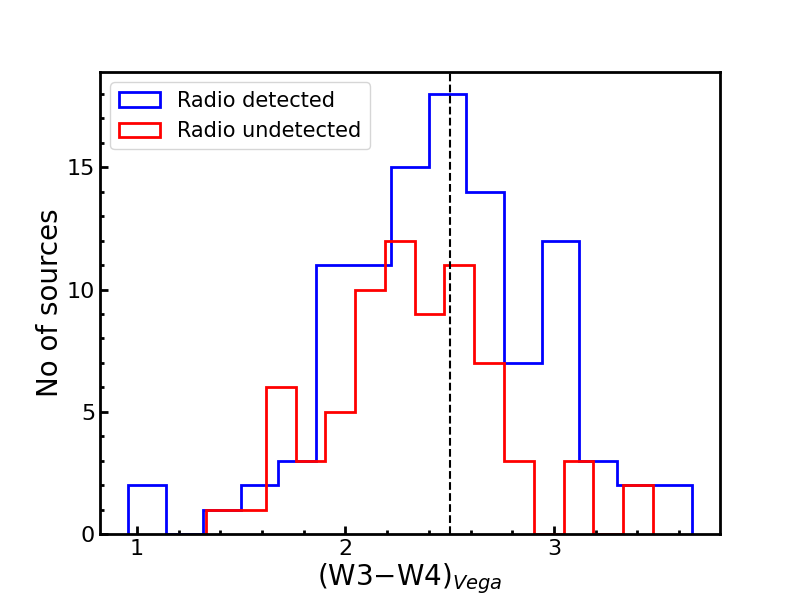}
    }
    \caption{Infrared colour-colour diagram for the sources with outflows in the radio-detected sample (left panel) and radio-undetected sample (middle panel). Black crosses refer to sources with two outflow components. The colour bar on the right indicates the total luminosity of the outflowing gas, and the vertical dashed line is the dividing line between AGN (left) and star-forming (right) according to \cite{2019A&A...622A..17S}. The right panel shows the distribution of W3$-$W4 colour. Here too, the vertical line at W3$-$W4 in 2.5 is the dividing line between AGN (left of the line) and star-forming (right of the line) sources \citep{2015MNRAS.451.1795C}.
    }
    \label{fig:w1-w2-w3-w4}
\end{figure*}

We generated a colour-colour diagram using W2$-$W3 and W1$-$W2 in the AB 
system for both the radio-detected and radio-undetected samples, and this is depicted in the left and middle panels of Fig. \ref{fig:w1-w2-w3-w4}. Sources in this
plot are classified into star formation and AGN, with a division at 
W2$-$W3 =0.8 According to this division \citep{2019A&A...622A..17S}, sources to 
the left are AGN dominated, and sources to the right are star formation dominated. 
Thus, in our radio-detected and radio-undetected sample, a large fraction of 
sources with outflows lie in the region occupied by star-forming galaxies.
Recently, \cite{2024RNAAS...8..188S} found that sources with $W2-W3$ $<$ 0.16 in the AB system
have very low specific star formation rate (sSFR) of $10^{-11.5}$ yr$^{-1}$.

We also investigated the W3$-$W4 colour, of our sample of sources with detected outflows, the distribution of which is shown in the right panel of Fig. \ref{fig:w1-w2-w3-w4} in the Vega system. Here, too, sources with W3$-$W4 $<$ 2.5 are AGN dominated, while sources with W3$-$W4 $>$ 2.5 are star formation dominated \citep{2015MNRAS.451.1795C}. The infrared colour-colour diagram and the W3$-$W4 colour indicate that a significant portion of sources with detected outflows falls within the region typically associated with star-forming galaxies, despite all our sources being classified as AGN based on the BPT diagram. This suggests that infrared colour is not a reliable metric for distinguishing between AGN-dominated and star-formation dominated sources. 

\begin{figure*}
\centering
    \hbox{
    \includegraphics[scale=0.45]{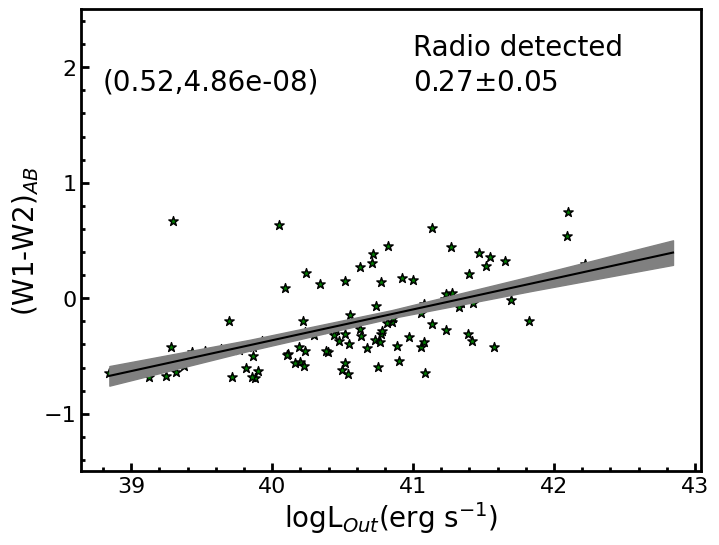}
    \includegraphics[scale=0.45]{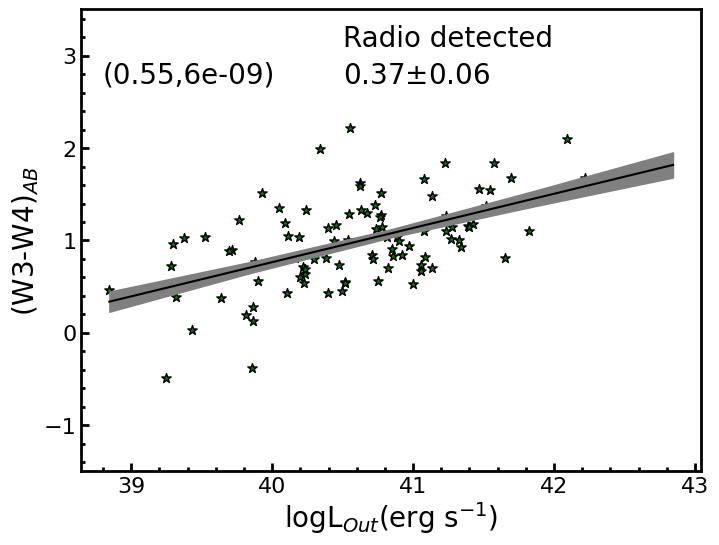}
    }
    \hbox{
     \includegraphics[scale=0.45]{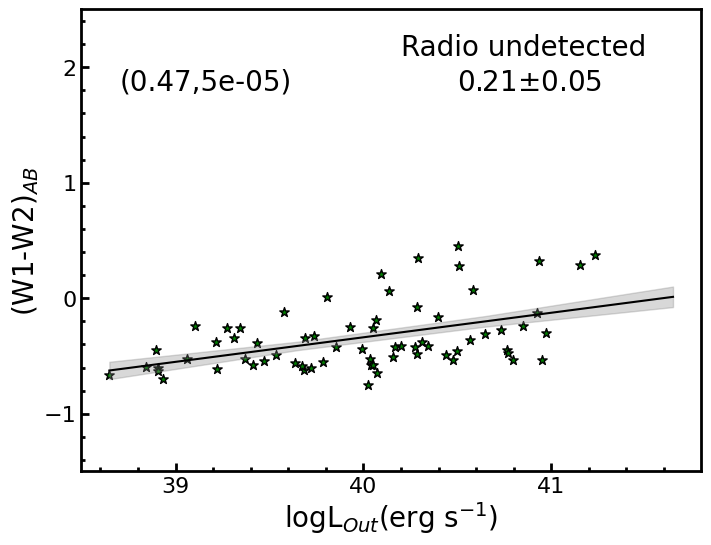}
    \includegraphics[scale=0.45]{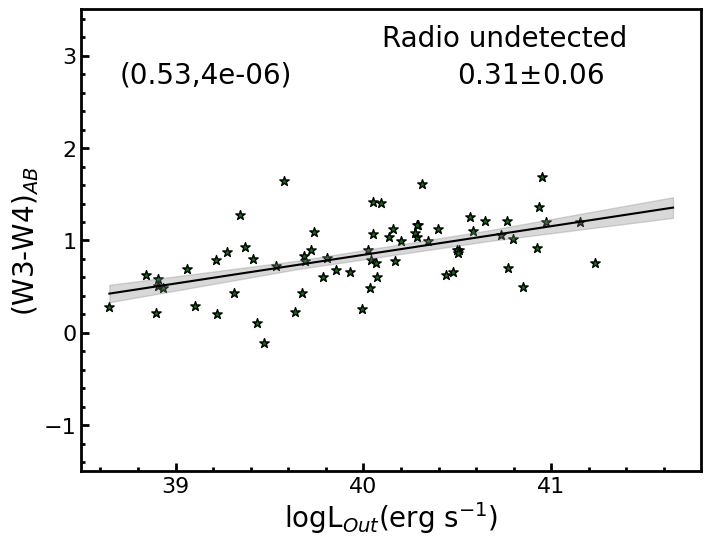}
    }
    \caption{Variation of infrared colours with outflow luminosity for radio-detected sample (upper panel) and for radio-undetected sample (lower panel). The correlation coefficient and p-value from the KS test are displayed in the upper-left corner of each plot, while the slope (see Section \ref{sec:outflow-AGN}) is indicated in the upper-right corner.}
    \label{fig:w1-w2-w3-w4-Lout}  
\end{figure*}

Our analysis indicates that sources exhibiting outflows tend to have redder infrared (IR) colours, as illustrated in Fig. \ref{fig:w1-w2-w3-w4}. Additionally, we find a positive correlation between the IR colours W1$-$W2 and W3$-$W4 and the luminosity of the outflowing component (Fig. \ref{fig:w1-w2-w3-w4-Lout}) in both cases. Interestingly, the correlation is stronger for the W3$-$W4 colour compared to W1$-$W2. This pattern is consistent across both the radio-detected and radio-undetected samples. These correlations suggest that dust in the vicinity of the outflows is likely the dominant contributor to the observed mid-infrared (MIR) emission.

Furthermore, the AGN in our sample, classified based on their BPT diagnostics, display increased redness in outflowing sources (see Fig. \ref{fig:w1-w2-w3-w4} and Fig. \ref{fig:w1-w2-w3-w4-Lout}), which can be attributed to polar dust scattering. Dust grains absorb ultraviolet (UV) and optical radiation, re-emitting it in the IR and thus producing the observed redder colours. This process not only affects the IR emission but also alters the ionization conditions of the surrounding gas, potentially influencing the chemical composition of the outflows and aiding the formation of various molecules \citep{2022A&A...658A..12J}.

Our results are in agreement with \cite{2013ApJ...768...22Z} who found that the MIR covering factor (the ratio of MIR luminosity to bolometric luminosity) correlates with the outflow component of [OIII]$\lambda$5007, with the correlation strengthening at longer wavelengths. This was interpreted as evidence for IR emission produced by dust embedded within the outflows. Observations of several Seyfert galaxies also reveal that a significant fraction of their MIR emission originates along their polar directions, extending from a few parsecs to several hundred parsecs from the central engine. This emission is likely due to dust in the narrow-line region and/or dust driven by outflows \citep{2013ApJ...771...87H, 2019MNRAS.484.3334S, 2024MNRAS.532.4645H}.

\subsection{Contribution of star formation to the outflows}\label{sec:sfr}
In the previous section, it was observed that the infrared properties of outflows 
reveal a redder colour similar to that of star-forming galaxies. This raises the 
need to assess the contribution of star formation to the outflows in our sample of sources, 
where outflows were detected. Though the sources are classified as AGN according to the 
BPT diagrams, the influence of nuclear star formation could still be 
present in them. We aimed to assess the role of star formation in influencing outflows
in the central 500 $\times$ 500 square pc region, which necessitates
investigation of the star formation characteristics. Numerous well-established 
tracers of star formation exist, such as strong emission lines in optical and 
infrared bands, as well as continuum emission from UV to radio 
wavelengths \citep{2012ARA&A..50..531K}. However, these tracers are often 
contaminated by AGN emissions. Recently, \cite{2018MNRAS.476..580S} demonstrated 
that the sSFR derived from the Balmer 4000 ~\AA~ break strength (D$_n$4000) is less 
impacted by AGN emission lines and thus can be a better diagnostic to constrain 
star formation \citep{2024MNRAS.527.7965W}. Consequently, we employed this method 
to examine the sSFR in our sample of sources.

We calculated D$_n$4000 by taking the ratio of the average of the flux density measurements 
in the blue spectral range (3525$-$3625 ~\AA~) to the average of the flux density measurements 
in the red spectral range (4150$-$4250 ~\AA~). This is defined as
\begin{equation}
D_n4000 = \frac{\int_{4150}^{4250} f_{\lambda} d\lambda / \int_{4150}^{4250} d\lambda} {\int_{3525}^{3625} f_{\lambda} d\lambda / \int_{3525}^{3625} d\lambda}
\end{equation}
The chosen spectral window is slightly 
different from the one originally defined by \cite{1983ApJ...273..105B}, however, 
captures the break cleanly and does not cover the metal absorption lines (see also \citealt{2024MNRAS.527.7965W} for the use of alternate
wavelength windows). This wavelength window covers both the Balmer limit of 3645~\AA~which is sensitive to young galaxies and the 4000~\AA~break. The lower bound of the blue region for the break is determined by taking into account the instrument's shortest wavelength coverage that corresponds to the redshifted wavelengths of all observed sources.

The value of D$_n$4000 parameter for our sample of radio-detected and radio-undetected sources ranges from 0.8 to 2.8. D$_n4000$ is close to unity for the galaxies dominated by O and B-type stars \citep{2024MNRAS.527.7965W}, whereas D$_n4000$ higher than 1.51 is for an old stellar population with age more than 1.1 Gyr \citep{2020A&A...633A..70P}.

A recent study by \cite{2020MNRAS.492...96B} on MaNGA sources found that regions of galaxies with  D$_n$4000 larger than 1.45 are 
quenched with very low star formation, though the exact values
of sSFR are not known but are less than 10$^{-11.5} yr^{-1}$. About 94\% of the sources in the radio-detected sample and 99\% of sources in the radio-undetected sample have D$_n4000$ larger than 1.45, which suggests substantially low or no star formation in this central region in our sample of sources. This may possibly be due to the negative feedback effect from AGN activity.

\begin{figure}
    \centering
    \hspace*{-1.0cm}\includegraphics[scale=0.5]{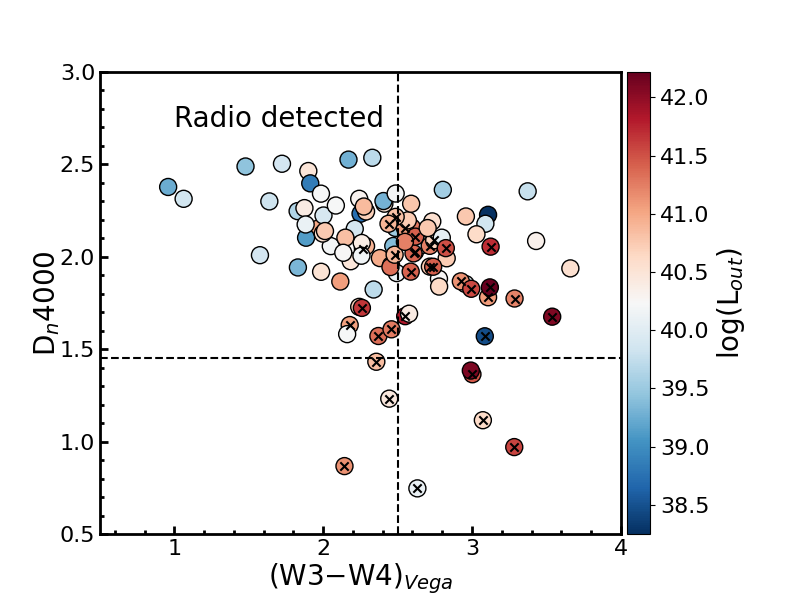}
    \hspace*{-1.0cm}\includegraphics[scale=0.5]{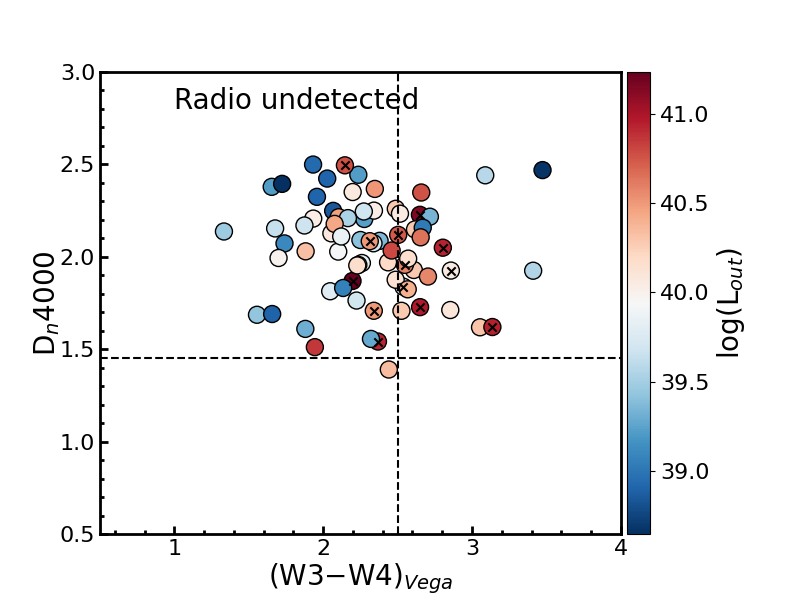}
    \caption{Position of the sources with outflows in W3$-$W4 vs D$_n$4000 plane for radio-detected (upper panel) and radio-undetected (lower panel) samples. The colour denotes the total luminosity of outflows. The black crosses are sources with two outflowing components. The vertical dashed line is $(W3-W4)_{Vega}=2.5$, the separation line between pure star-forming sources and AGN. The horizontal line is for $D_n4000=1.45$. }
    \label{fig:W3-W4-D4000}
\end{figure}

\subsection{Origin of outflows: AGN v/s star formation}\label{sec:origin-outflow}
 In Sections \ref{sec:IR}, we observed that infrared diagnostics alone are insufficient to distinguish whether strong outflows originate from AGN activity or purely from star formation. However, in Section \ref{sec:sfr}, using optical diagnostics such as the Balmer break, we found that in sources with outflows, the star
formation is very low or negligible. By combining these two diagnostic methods and analyzing the position of the sources with outflows in the D$_n4000$ v/s W3$-$W4 plane, it would be
possible to identify if the detected outflows are due to star formation and/or AGN activity. We show in Fig. \ref{fig:W3-W4-D4000} the infrared colour versus the Balmer break plot. From this figure, it is evident that most of our sources are situated in the AGN-dominated region. This new diagnostic diagram clearly indicates that the outflows found in sources with and without radio emission are due to processes related to AGN. About 5\% of the radio-detected sources with outflows lie in the region occupied by star formation with redder colours. In all these sources, both blueshifted (approaching component of outflows) and red-shifted (receding component of outflows) were detected. The redshifted component of the outflow, being
located below the plane of the galaxy, is likely to be obscured by dust, and the observations of such sources to be redder in colour
is not unexpected \citep{2024Natur.630...54B}. Irrespective of that, the contribution from both AGN and star formation to the observed outflows in these minority of sources could not be ruled out. Also, a large fraction of sources with D$_n4000 > 1.45$ (thus negligible star formation) have redder colours, and this is likely due to the interaction of the outflowing gas with dust \citep{2022A&A...658A..12J}. 

\subsection{Cause of radio emission}\label{sec:radio-sfr}
\begin{figure}
\centering
      \hspace*{-0.2cm}\includegraphics[scale=0.45]{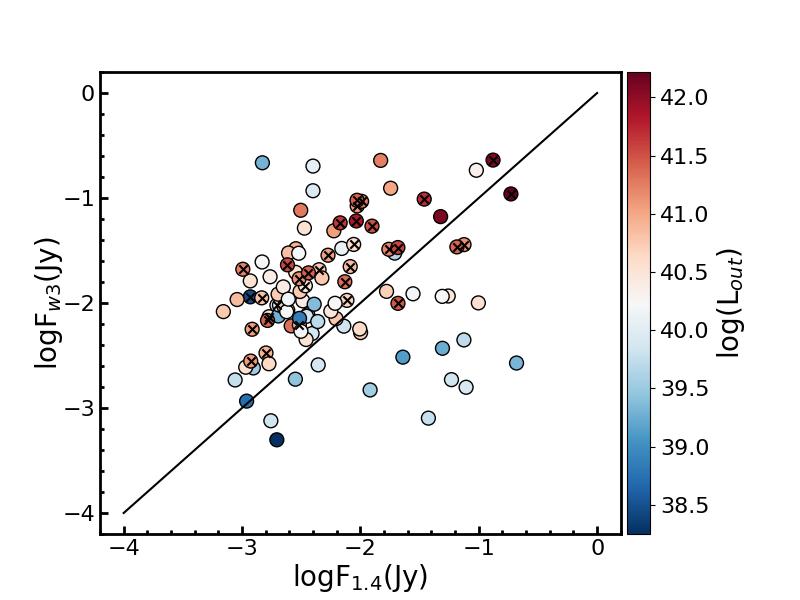}
\caption{Location of the sources with outflows in the radio-detected sample in the F(W3) v/s F(1.4GHz) plane. The solid black solid line is the F$_{W3}$ = F$_{1.4}$ line. The black crosses are sources with two components of outflows. The colour denotes the luminosity of outflows.}
\label{fig:w3-F14}
\end{figure}

\begin{figure}
    \centering
    \hspace*{-0.4cm}\includegraphics[scale=0.5]{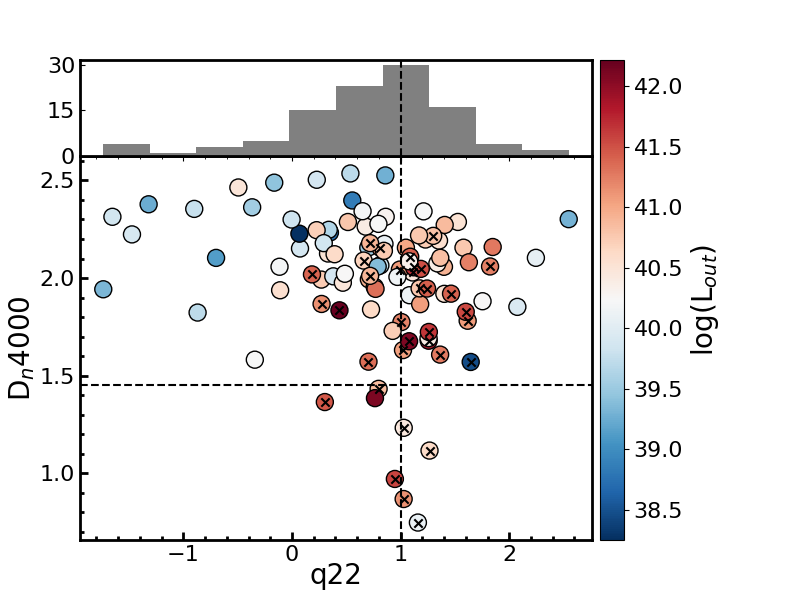}
    \caption{Position of outflow detected sources in the radio-detected sample in q22 vs D$_n$4000 plane. The vertical line is $q22=1.0$, and the horizontal dashed line is $D_n4000=1.45$.  }
    \label{fig:q22-4000}
\end{figure}

From Section \ref{sec:origin-outflow}, it is clear that the observed outflows are due to AGN in both the samples of radio-detected and radio-undetected sources. Therefore, the observed radio emission in our radio-detected sample is unlikely to be due to star formation activities in their host galaxies, however, attributed to processes related to AGN such as the presence of low power radio jets, accretion disk corona as well as shocks due to outflows \citep{2019NatAs...3..387P, 2024MNRAS.528.3696L}. In this section, we aim to understand the origin of radio emission in our radio-detected sample making use of diagnostic plots available in the literature. We show in Fig. \ref{fig:w3-F14} the location of sources with
outflows in the radio-detected sample in the F$_{W3}$ v/s F$_{1.4}$ GHz plane. For this plot, the radio flux density values were taken from the FIRST survey, while the flux density corresponding to the W3 band of WISE was taken from the WISE catalogue. Also shown in the same plot is the $F_{W3} = F_{1.4}$ line. According to  \cite{2021ApJ...910...64K} sources below the line are radio AGN, while those above the line are starburst dominated AGN. We also checked the q22 parameter defined as \begin{equation}
q22 = log\left(F_{22} / F_{1.4} \right )
\end{equation}
Here, F$_{22}$ and F$_{1.4}$ are the flux densities in the W4 band 
of the WISE and 1.4 GHz from FIRST respectively. The histogram of the q22 parameter is shown in Fig. \ref{fig:q22-4000}. Here, too, about 40\% of the sources have q22 greater than unity, favouring star formation processes to be the cause of radio emission in them. In summary, although the q22 parameter and the F$_{W3}$ v/s F1.4 diagnostics indicate that in a large fraction of the sources, the observed radio emission is likely to be associated with star formation activity, the plot of
D$_n4000$ against q22 (Fig. \ref{fig:q22-4000}) shows that all sources barring six sources, lie in the AGN dominated region. This reddening could be due to dust scattering of AGN radiation. This reinforces that the observed radio emission in our radio-detected sample is indeed AGN dominated. High resolution radio observations are the only direct way to identify which among the processes related to AGN are the cause of the observed radio emission in our sample. Though the detection of core jet structure is an unambiguous evidence of jet that produces the observed radio emission, the resolution of FIRST images used
in this work is insufficient. In the absence of this, in all further discussion, we assume that the observed radio emission is due to jet emission.

\subsection{Correlation of outflow properties with physical properties of AGN}\label{sec:outflow-AGN}
From various diagnostics, it is clear that the detected outflows
are due to AGN.  In this scenario, the driving force of outflows could be 
either from the radiation energy or the radio jets from AGN.  To explore this, 
we analyzed the outflow properties alongside AGN properties such as the M$_{BH}$, bolometric luminosity ($L_{Bol}$)and Eddington 
ratio ($\lambda_{Edd}$) for both the radio-detected and radio-undetected samples 
as well as the radio jet power (P$_{Jet}$) for the radio-detected sample.
We determined M$_{BH}$ adopting the dynamical method, 
using the M$_{BH}$ $-$ $\sigma_{\star}$ relation, where $\sigma_{\star}$ represents 
the stellar velocity dispersion. The $\sigma_{\star}$ values were obtained from 
the \textit{Pipe3d} catalogue \citep{2022ApJS..262...36S}, derived through stellar 
synthesis population modelling within one effective radius. Following the 
relation provided in \cite{2019MNRAS.487.3404B}, which has been validated for 
both Type 1 and Type 2 AGN, by \cite{2019MNRAS.487.3404B}, we calculated M$_{BH}$ 
for all the sources in our sample.

We calculated L$_{Bol}$ from H$\alpha$ luminosity by following \cite{2007ApJ...670...92G} and \cite{Greene_2005} wherein the luminosity of the H$\alpha$ line was determined from the H$\alpha$ flux taken from DAP and corrected for extinction effect.
We also calculated $\lambda_{Edd}$, from Eddington luminosity L$_{Edd}$, the maximum luminosity emitted if the source is in hydrodynamical equilibrium and L$_{Bol}$, as the ratio of L$_{Bol}$ to L$_{Edd}$. The L$_{Edd}$ defined as
\begin{equation}
L_{Edd} = 1.26 \times 10^{38}\left(\frac{M_{BH}} {M_{\odot}}\right) erg\:s^{-1}
\end{equation}

\begin{figure}
    \centering
    \hspace*{-0.4cm}\includegraphics[scale=0.45]{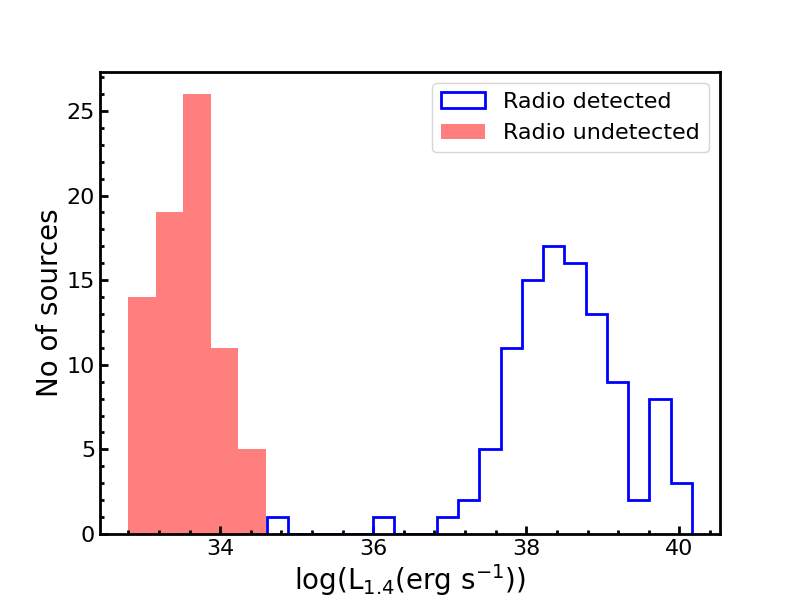}
    \caption{Distribution of L$_{1.4}$ for the radio detetected sample (blue color). The shaded red region shows the upper limit of L$_{1.4}$ for the radio-undetected by considering the detection limit of the FIRST survey, which is 0.5 mJy.}
    \label{fig:L14}
\end{figure}

For the sample of radio-detected sources, we estimated P$_{Jet}$ by following \cite{2010ApJ...720.1066C} and considering that the radio emission in these sources is jet emission. For this, we used the 1.4 GHz luminosity (L$_{1.4}$) calculated using the integrated flux densities from FIRST survey and corrected for 
redshift effect assuming a spectral index ($S_{\nu} \propto\nu^{-\alpha}$) of $0.7$ 
\citep{2002AJ....124..675C}. The distribution of L$_{1.4}$ for our sample of radio-detected and radio-undetected sources is shown in Fig. \ref{fig:L14}. For radio-undetected sources, the L$_{1.4}$ values are the upper limits that were calculated using the detection limit of the FIRST survey, which is 0.5 mJy. 

After calculating these physical parameters of AGN, we compared them with outflow 
properties to explore potential correlations. In cases where two outflowing components were detected, for the total outflow rate or the total kinetic power of outflows, we used the sum of the quantities deduced from both components. We performed a statistical linear 
correlation test, the Pearson test, to identify significant 
correlations in terms of correlation coefficients and p-values. For parameters exhibiting significant correlations, we employed the Bayesian linear regression method using the \textit{LINMIX$\_$ERR} \citep{2007ApJ...665.1489K} to fit a power-law relationship between the variables in log-log space. This method takes account of errors in both axes. If the AGN parameter is X$_{AGN}$ and outflow parameter is Y$_{out}$ then the fitted function has the form

\begin{equation}
Y_{out} = A (X_{AGN})^{\alpha}
\end{equation}
or 
\begin{equation}
log(Y_{out}) = log A + \alpha log(X_{AGN})
\end{equation}

where \( A \) is the multiplication constant and \( \alpha \) is the power-law exponent. The best-fit values for \( \alpha \) and the correlation coefficients for different parameters are summarized in Table \ref{tab:para}.

\begin{table*}
\centering
\caption{Results of the fits to the observed data. Here, R and p are the correlation coefficient and probability for no correlation, respectively for the Pearson correlation test. The quoted values of $\alpha$, the power law exponent are the mean and the one standard deviation (1$\sigma$).}
\begin{tabular}{lrrrr}
\hline
Parameter & \multicolumn{2}{c}{Radio-detected} & \multicolumn{2}{c}{Radio-undetected} \\
Y$_{out}$ v/s X$_{AGN}$  &  (R,p)    & $\alpha$            &  (R,p)  & $\alpha$          \\ 
\hline
\hspace{0.02cm}\\
KP$_{out}$ v/s L$_{Bol}$ &  (0.72, 7$\times 10^{-18}$) & 1.18$\pm$0.14 & (0.52, 2$\times10^{-6}$) & 0.81$\pm$0.20  \\

$\dot{M}_{out}$ v/s L$_{Bol}$ &  (0.74, 3$\times 10^{-19}$) & 1.07$\pm$0.12 & (0.60, 9$\times10^{-9}$) & 0.97$\pm$0.17 \\

KP$_{out}$ v/s P$_{Jet}$ &  (0.36, 2$\times10^{-4}$) & 0.56$\pm$0.14 & -- & -- \\

$\dot{M}_{out}$ v/s P$_{Jet}$ &  (0.28, 0.004) & 0.43$\pm$0.12 & -- & -- \\

KP$_{out}$ v/s P$_{Jet}$ [log($\frac{L_{Bol}}{P_{Jet}}$)$>0.4$] &  (0.64, 1$\times10^{-11}$) & 0.99$\pm$0.14 & -- & -- \\

$\dot{M}_{out}$ v/s P$_{Jet}$ [log($\frac{L_{Bol}}{P_{Jet}}$)$>0.4$] &  (0.58, 5$\times 10^{-9}$) & 0.79$\pm$0.12 & -- & -- \\

\hspace{0.02cm}\\
\hline 
\end{tabular}
\label{tab:para}
\end{table*}

\begin{figure*}
    \centering
    \vbox{
    \hbox{
\hspace*{-0.2cm} \includegraphics[scale=0.48]{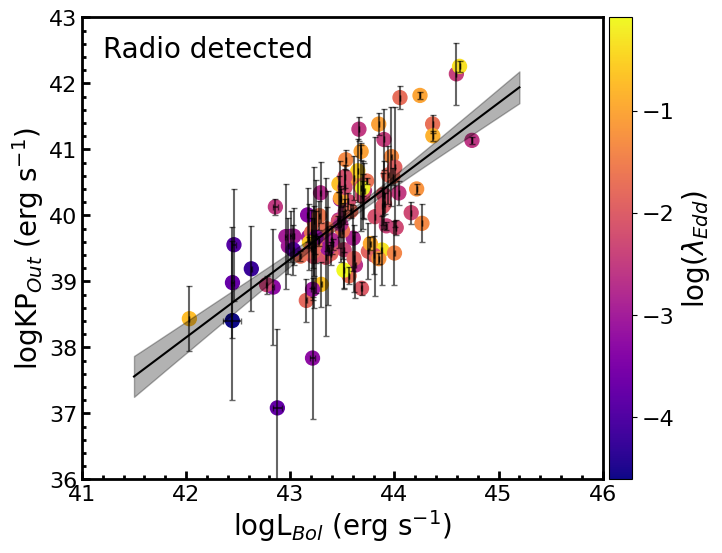}
\hspace*{-0.2cm}\includegraphics[scale=0.48]{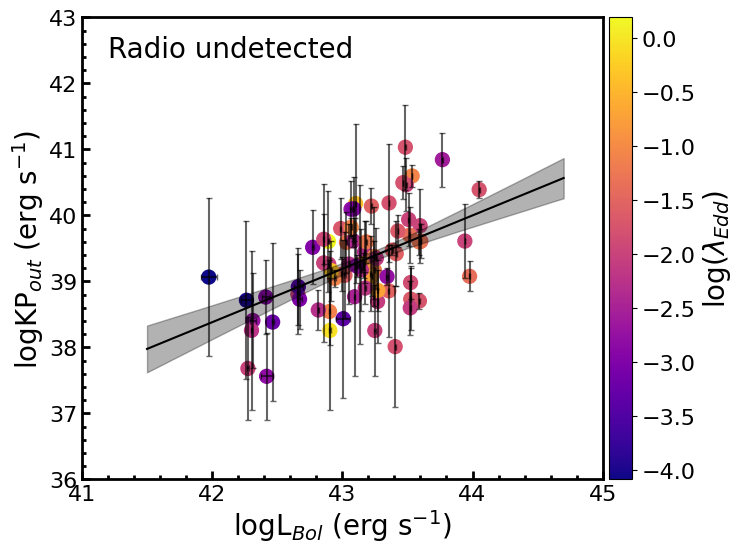}
 }
\hbox{
\hspace*{-0.2cm} \includegraphics[scale=0.48]{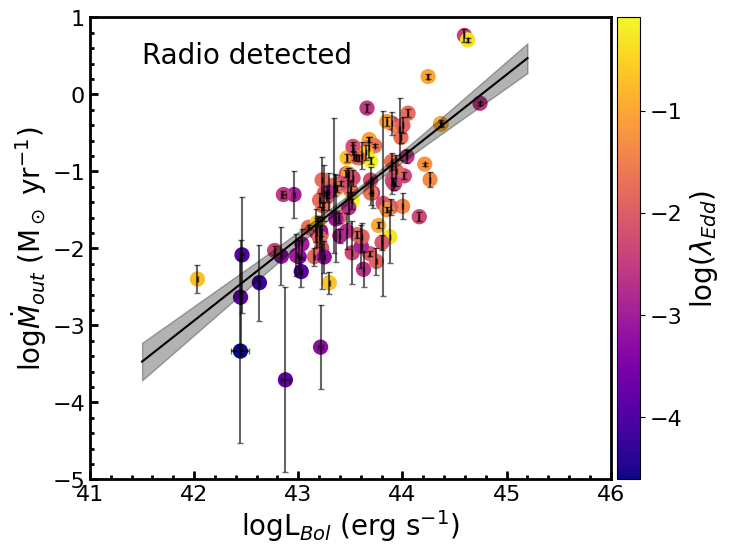}
\hspace*{-0.2cm}\includegraphics[scale=0.48]{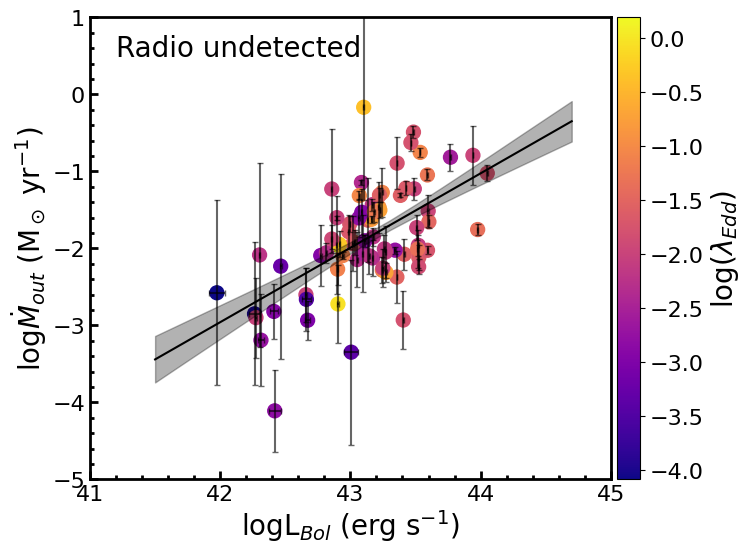}
 }
 }
    \caption{Upper panel: Variation of total kinetic power of warm ionized outflow with L$_{Bol}$  for radio$-$detected (left panel) and radio$-$undetected sources (right panel). Lower panel: Variation of total outflow rate with L$_{Bol}$ for the radio-detected sources (left panel) and radio-undetected sources (right panel). In each plot, the scatter points represent our data with 1$\sigma$ error bars, while the solid line and shaded region indicate the fitted line with a 1$\sigma$ confidence band. The color in each plot corresponds to $\lambda_{Edd}$.}
\label{fig:agn-bol}
\end{figure*}

From Table \ref{tab:para}, it is evident that both \(\dot{M}_{out}\) and \(KP_{out}\) are significantly correlated with \(L_{Bol}\) for both radio-detected and radio-undetected sources. This finding aligns with previous studies in the literature \citep{2017A&A...601A.143F, 2019A&A...628A.118B, 2023A&A...679A..84M}. While our results confirm the correlation between \(\dot{M}_{out}\) and \(L_{Bol}\) noted in earlier works, we also emphasize the differences in the correlations between the two samples. Notably, the higher correlation coefficient and lower p-values for radio-detected sources suggest that this correlation is stronger in radio-detected sources than in their radio-undetected counterparts.

Examining the relationship of these outflow parameters with AGN luminosity, we found that for the radio-detected sample, we observed \(\dot{M}_{out} \propto L_{Bol}^{1.07\pm0.12}\). In contrast, for the radio-undetected sample, \(\dot{M}_{out} \propto L_{Bol}^{0.97\pm0.17}\). Though the slopes are consistent within 1\(\sigma\), \(KP_{out}\) demonstrates a more pronounced difference: we found \(KP_{out} \propto L_{Bol}^{1.18\pm0.14}\) for radio-detected sources, while for the radio-undetected sample, we found \(KP_{out} \propto L_{Bol}^{0.81\pm0.20}\). This indicates a steeper slope by 1\(\sigma\) for the radio-detected sources compared to their undetected counterparts. This trend is illustrated in Fig. \ref{fig:agn-bol}, where we plot \(L_{Bol}\) against \(KP_{out}\) and \(\dot{M}_{out}\), color-coded by \(\lambda_{Edd}\). In the radio-detected category of sources, those with larger $\lambda_{Edd}$ preferentially occupy the region with larger
KP$_{out}$. This suggests a relationship between outflow power and \(\lambda_{Edd}\) for these sources. Conversely, this trend is not as clear for radio-undetected sources, as depicted on the right side of Fig. \ref{fig:agn-bol}.

The correlations observed in Fig. \ref{fig:agn-bol}, suggest multiple mechanisms are at play in driving outflows for radio-detected sources. While radiation from AGN is likely the primary driver of outflows in both radio-detected and radio-undetected sources, radio jets may serve as an additional mechanism that enhances outflow kinematics in radio-detected sources. This could explain the steeper correlation between \(KP_{out}\) and \(L_{Bol}\) in the radio-detected sample and the stronger correlation of \(KP_{out}\) with \(\lambda_{Edd}\). Although there is a general upward trend of outflow properties with \(L_{Bol}\), the scatter in the plots (see Fig. \ref{fig:agn-bol}) may be attributed to the complex interplay between outflows and the quantity or geometry of dense gas in the nuclear regions of these sources \citep{2022A&A...658A.155R}. Moreover, the colour coding in Fig. \ref{fig:agn-bol} indicates that sources with high \(\lambda_{Edd}\) tend to have elevated values of \(L_{Bol}\), \(\dot{M}_{out}\), and \(KP_{out}\). 
This interpretation highlights the nuanced role of AGN radiation and radio jets in influencing outflow characteristics, suggesting a more complex feedback mechanism that merits further investigation.

\begin{figure}
    \centering
    \hspace*{-0.8cm}\includegraphics[scale=0.43]{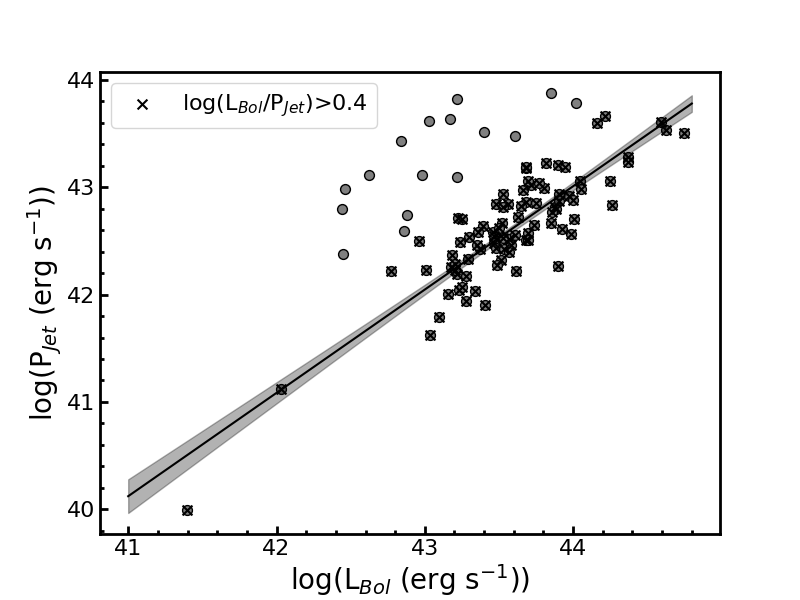}
    \hspace*{-0.5cm}\includegraphics[scale=0.48]{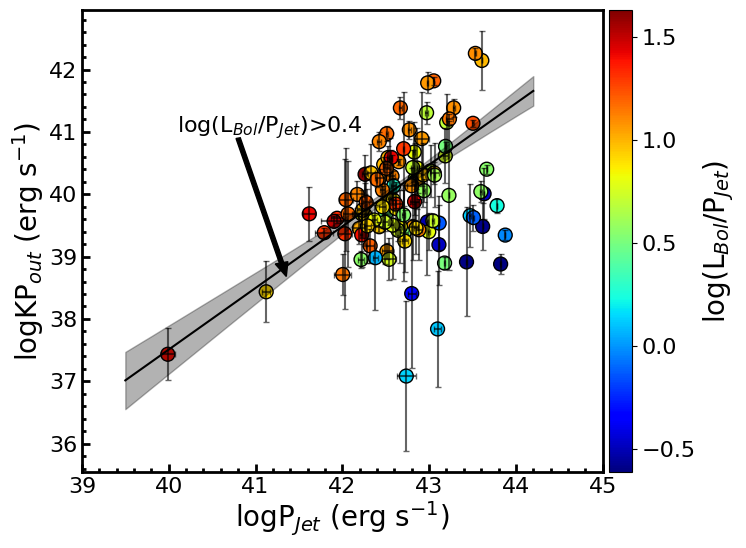}
    \caption{Upper panel: Jet power vs bolometric luminosity. The sources with log($\frac{L_{Bol}}{P_{Jet}}$)$>0.4$ are shown with crosses. The fitted line is for the sources with log($\frac{L_{Bol}}{P_{Jet}}$)$>0.4$. Lower panel: The variation of kinetic power of outflow with jet power. The solid line with the shaded region is the fitted line with 1$\sigma$ uncertainty for the given condition. The colour in each plot corresponds to $\frac{L_{Bol}}{P_{Jet}}$ in the log scale.}
    \label{fig:jet-out}
\end{figure}

For the radio-detected sample of sources, we found a flat relation between the outflow properties and jet power. We obtained the best fit scaling relation of KP$_{out} \propto  P_{Jet}^{0.56\pm0.14}$ and $\dot{M}_{out} \propto  P_{Jet}^{0.43\pm0.12}$. The results of the fits are given in Table \ref{tab:para}. We also found the ratio of log($\frac{L_{Bol}}{P_{Jet}}$) to have a bimodal behaviour with a dividing limit at 0.4. We noticed beyond this limit, P$_{Jet}$ is very strongly correlated with L$_{Bol}$ with a slope of $0.96\pm0.06$  which can be seen in the upper panel of Fig. \ref{fig:jet-out}.  In the lower panel of Fig. \ref{fig:jet-out}, we show the correlation between KP$_{out}$ and P$_{Jet}$. Here, the sources are colour coded with log($\frac{L_{Bol}}{P_{Jet}}$). We found that beyond the limit of log($\frac{L_{Bol}}{P_{Jet}}$)=0.4 i.e. for log($\frac{L_{Bol}}{P_{Jet}}$)$>0.4$ the correlation between KP$_{out}$ and P$_{Jet}$ is significantly strong and steep with KP$_{out} \propto  P_{Jet}^{0.99\pm0.14}$. Below this limit, i.e. for higher P$_{Jet}$ with similar L$_{Bol}$, KP$_{out}$ is lower. This may be due to high power jet encountering lower interaction with the cloud where [OIII]$\lambda5007$ originates, reducing the outflow luminosity and leading to low outflow characteristics such as KP$_{out}$ and $\dot{M}_{out}$. Alternatively, a high-power jet can ionize the gas to its higher ionization state, leading to low luminous outflow. Thus at any jet power, significant dominance of the jet power over bolometric luminosity can lead to weaker outflows.

This interpretation highlights the nuanced role of AGN radiation and radio jets in shaping outflow characteristics, suggesting a more complex feedback mechanism that warrants deeper exploration. Future studies can leverage high-resolution, multi-wavelength observations to clarify the relative contributions of radiation-driven and jet-driven feedback processes. Spatially resolved spectroscopic studies, combined with detailed radio imaging, can help establish a clearer link between jet morphology and outflow kinematics. Additionally, theoretical modelling and simulations can provide further insights into the interplay between AGN-driven winds, jets, and the surrounding ISM. Investigating these aspects across a broader range of AGN types and host environments will be crucial for building a more comprehensive understanding of AGN feedback and its impact on galaxy evolution.

\section{Summary}\label{sec:summary}

In this study, we conducted a comparative analysis of outflow properties in radio-detected and radio-undetected sources in their central region of 500$\times$500 square pc. Our total sample consists of 538 AGN with detected [OIII]$\lambda$5007 line, mainly Seyfert and LINER type, of which 197 are radio-detected and 341 are radio-undetected. The objective was to identify outflows, study their properties, and constrain the role of AGN radiation and/or jets in driving outflows. For this, we used spatially resolved optical spectroscopic data from MaNGA and radio observations from FIRST surveys. We studied the properties of outflows and then compared them with AGN properties. Additionally, we explored the relationship between radio properties and outflow characteristics within the radio-detected sources. We summarize our main
findings below. 

\begin{enumerate}

\item To detect outflows, we carried out multiple Gaussian fits to the observed [OIII]$\lambda$5007 line. In the radio-detected sample, 56$\pm$7\% of sources showed evidence of outflows. However, in the radio-undetected sample, 25$\pm$3\% of sources showed
    outflows. Thus, in our sample, the outflow detection rate is higher in radio-detected sources compared to radio-undetected sources.

\item On separating our sample of sources into Seyferts and LINERs, outflows are detected more in Seyferts (66$\pm$7\%) relative to LINERs (15$\pm$2\%). This is true for both the radio-detected and radio-undetected samples. The mass outflow rate and outflow power are higher for Seyferts than LINERs, but the velocity structures are higher for LINERs compared to Seyferts. 

\item In both the radio-detected and radio-undetected sample, for a majority of
     sources ($\sim$ 80\%), we found the [OIII]$\lambda$5007 line to have a blue asymmetry in addition
     to the narrow component. Also,  in a minority of sources, in addition to the
     blueshifted component, we also observed the redshifted component. The
     blueshifted component could be the approaching side of the outflow located
     above the plane of the galaxy, and the redshifted component could be the
     receding side of the outflow, located below the plane of the galaxy.

\item We observed distinct differences in the kinematics of the outflowing gas between the radio-detected and radio-undetected samples. The radio-detected sources exhibit higher velocity, larger velocity dispersion, greater asymmetry, larger outflow mass, and stronger kinematic power compared to the radio-undetected sources.

\item We found that in the infrared bands, more luminous outflows appear redder in colour compared to weaker outflows. Infrared colours show a positive correlation with outflow luminosity, with this dependence becoming more pronounced in the mid-infrared band. This trend is primarily attributed to the presence of larger amounts of polar dust in the more powerful outflows.

\item  We found a strong correlation between the outflow characteristics, such as
    the $\dot{M}_{out}$ and KP$_{out}$ of the outflow with the L$_{Bol}$.  Such a correlation points to radiation from AGN being the primary
    driver for outflows in both radio-detected and radio-undetected samples.
    However, this correlation between the outflow characteristics with the
    bolometric luminosity is mildly steeper for the radio-detected sample compared to the radio-undetected sample.  This suggests that in the radio-detected sample, radio jets could play an additional modest secondary role over and above the dominant role played by radiation in enhancing outflow kinematics.

\item Outflow characteristics are also found to show a correlation with
    $\lambda_{Edd}$. Sources with higher $\lambda_{Edd}$ appear to have higher
    L$_{Bol}$, M$_{out}$ and KP$_{out}$. This is true for both the radio-detected and
    radio-undetected samples.

\item For the radio-detected sample, we observed a bi-modality in
    the distribution of log ($\frac{L_{Bol}}{P_{Jet}}$), with the dividing line
    at log($\frac{L_{Bol}}{P_{Jet}}$) = 0.4. In the correlation between the
    kinetic power of outflows and jet power, we found that at any jet power,
    significant dominance of the jet power over the bolometric luminosity can
    lead to weaker outflows.

\item We found the value of the D$_n4000$ parameter for our sample of radio-detected and radio-undetected sources to range between 0.8 and 2.8. About 94\% of the sources in the radio-detected sample and 99\% of sources in the radio-undetected sample have D$_n4000$ larger than 1.45, pointing to sSFR lesser than 10$^{-11.5}$ yr$^{-1}$, which may possibly be due to negative AGN feedback.

\end{enumerate}

Our findings suggest that ionized gas outflows, driven by the interaction between AGN radiation/winds and the ISM, are common across all AGN. However, the presence of radio jets appears to affect gas kinematics further, leading to a higher rate of outflow detection in radio-detected sources, as evidenced by our study. Further investigations using high-resolution, multi-wavelength observations for different types of AGN may provide more insights towards understanding the feedback processes 
of radiation/winds and jets in greater detail. For example, high-resolution radio observations along with observations at other wavelengths would allow for a detailed spatial correlation between the morphology of radio jets and the outﬂowing gas, and understanding their kinematics.



\section{Acknowledgments}
\begin{acknowledgments}
We thank the reviewer for their comments and suggestions, which have helped to improve the manuscript. We also acknowledge Dr. Chris Harrison, at Newcastle University for useful suggestions in the initial stage.
This publication uses data from the MaNGA (Mapping Nearby Galaxies at APO) survey, which is one of the Sloan Ditial Sky Survey (SDSS) IV programs. Funding for the Sloan Digital Sky Survey IV has been provided by the 
Alfred P. Sloan Foundation, the U.S. Department of Energy Office of Science, and the Participating Institutions. 

SDSS-IV acknowledges support and resources from the Center for High Performance Computing at the University of Utah. The SDSS website is www.sdss4.org.

SDSS-IV is managed by the Astrophysical Research Consortium for the Participating Institutions of the SDSS Collaboration including the Brazilian Participation Group, the Carnegie Institution for Science, Carnegie Mellon University, Center for Astrophysics | Harvard \& Smithsonian, the Chilean Participation Group, the French Participation Group, Instituto de Astrof\'isica de Canarias, The Johns Hopkins University, Kavli Institute for the Physics and Mathematics of the Universe (IPMU) / University of Tokyo, the Korean Participation Group, Lawrence Berkeley National Laboratory, Leibniz Institut f\"ur Astrophysik Potsdam (AIP),  Max-Planck-Institut f\"ur Astronomie (MPIA Heidelberg), Max-Planck-Institut f\"ur Astrophysik (MPA Garching), Max-Planck-Institut f\"ur Extraterrestrische Physik (MPE), National Astronomical Observatories of China, New Mexico State University, New York University, University of Notre Dame, Observat\'ario Nacional / MCTI, The Ohio State University, Pennsylvania State University, Shanghai Astronomical Observatory, United Kingdom Participation Group, Universidad Nacional Aut\'onoma de M\'exico, University of Arizona, University of Colorado Boulder, University of Oxford, University of Portsmouth, University of Utah, University of Virginia, University of Washington, University of Wisconsin, Vanderbilt University, and Yale University. This project makes use of the MaNGA-Pipe3D dataproducts. We thank the IA-UNAM MaNGA team for creating this catalogue, and the Conacyt Project CB-285080 for supporting them

This publication uses radio observations carried out using the National Radio Astronomy Observatory facilities Very Large Array (VLA) of FIRST survey. The National Radio Astronomy Observatory is a facility of the National Science Foundation operated under a cooperative agreement by Associated Univerties, Inc. This work has made use of the NASA Astrophysics Data System (ADS)\footnote{https://ui.adsabs.harvard.edu/} and the NASA/IPAC extragalactic database (NED)\footnote{https://ned.ipac.caltech.edu}. 
PN thanks the Council of Scientific and Industrial Research (CSIR), Government of India, for supporting her research under the CSIR Junior/Senior research fellowship program through the grant no. $09/079(2867)/2021-EMR-I$.
\end{acknowledgments}

%
\pagebreak
\vspace{15cm}
\facilities{SDSS, MaNGA, FIRST}


\software{
Topcat \citep{2005ASPC..347...29T}, Numpy \citep{harris2020array}, 
Astropy \citep{2022ApJ...935..167A}, 
Scipy \citep{2020SciPy-NMeth}, Matplotlib \citep{Hunter:2007}, PyNeb \citep{2015A&A...573A..42L}, LINMIX$\_$ERR \citep{2007ApJ...665.1489K}
}


\pagebreak

\appendix

\restartappendixnumbering 

\section{Sample and emission line fits}\label{app-bpt-samp-line}
\begin{figure}[h!]
\centering
\hbox{
\includegraphics[scale=0.48]{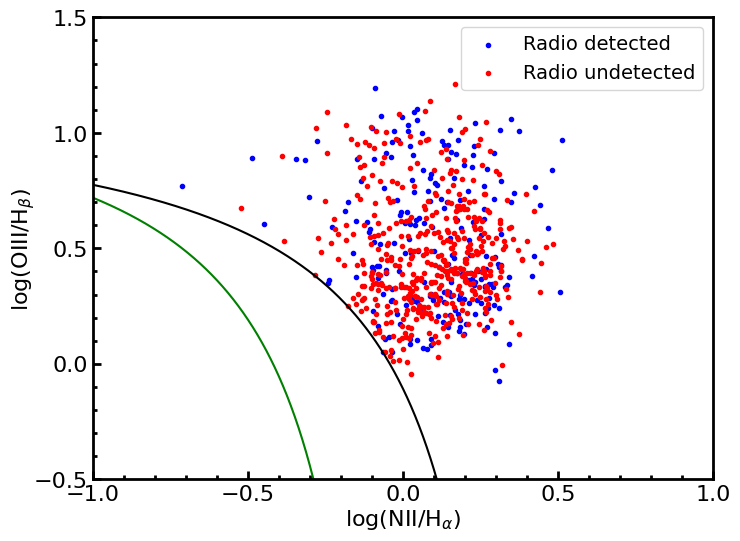}
\includegraphics[scale=0.48]{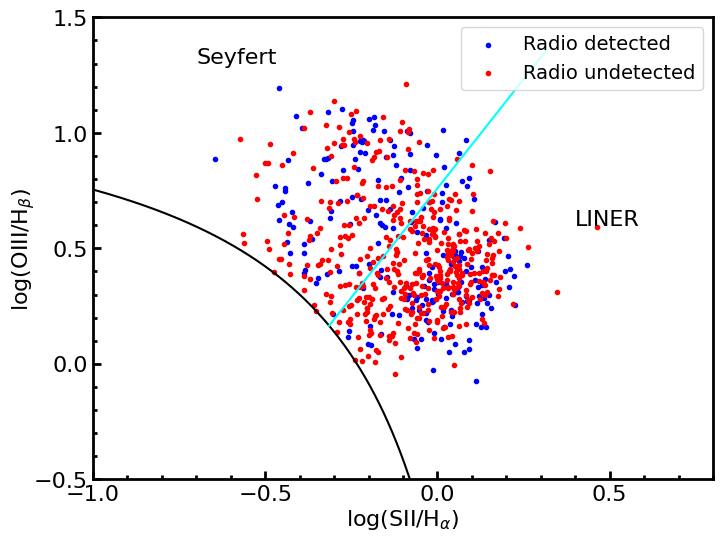}
}
\caption{Positions of the sources in the [NII]/H$\alpha$ BPT diagram (left) and [SII]/H$\alpha$ BPT diagram (right). The black and green solid lines separate the region occupied by AGN and star-forming galaxies according to \protect\cite{2001ApJS..132...37K} and \protect\cite{2003MNRAS.346.1055K}, respectively. The cyan solid line separates Seyfert galaxies and LINERs \protect\citep{2001ApJS..132...37K}. Filled blue and red circles refer to radio-detected and radio-undetected sources.}
\label{fig:bpt}
\end{figure}

As described in Section \ref{sec:samp}, our sample to study the mechanisms that trigger warm ionised outflows in AGN, is from MaNGa,
which contains both AGN and non-AGN sources. We, therefore selected the AGN sources by constructing BPT diagrams. The positions of the
sources in the final sample in the BPT diagrams are shown in Fig. \ref{fig:bpt}. The AGN sources thus selected were divided into radio-detected and radio-undetected ones.

\begin{figure}[h!]
    \centering
    \hspace*{-0.7cm}\includegraphics[scale=0.50]{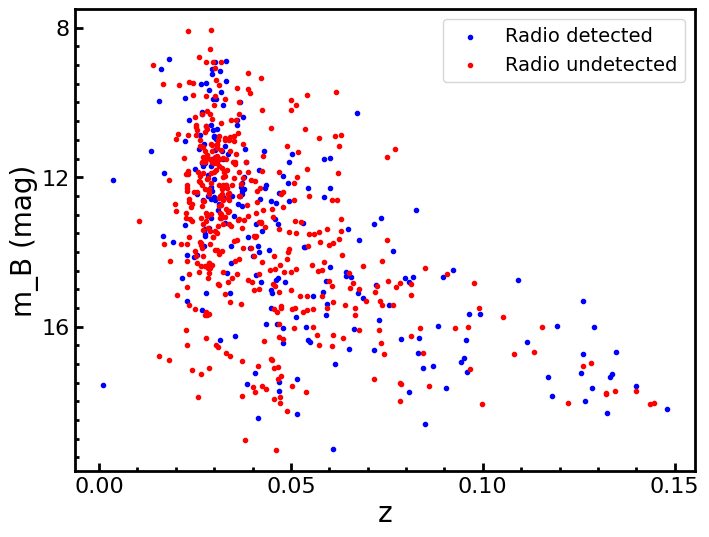}
    \caption{Distribution of sources in the redshift v$/$s  B-band apparent magnitude plane. Here, the filled blue and red circles refer to the radio-detected and radio-undetected sources, respectively.}
    \label{fig:lum-z}
\end{figure}

The distribution of the sources in the redshift and optical B-band brightness plane is shown in Fig. \ref{fig:lum-z}. 

\begin{figure}[h!]
    \centering
    \hbox{
    \includegraphics[scale=0.3]{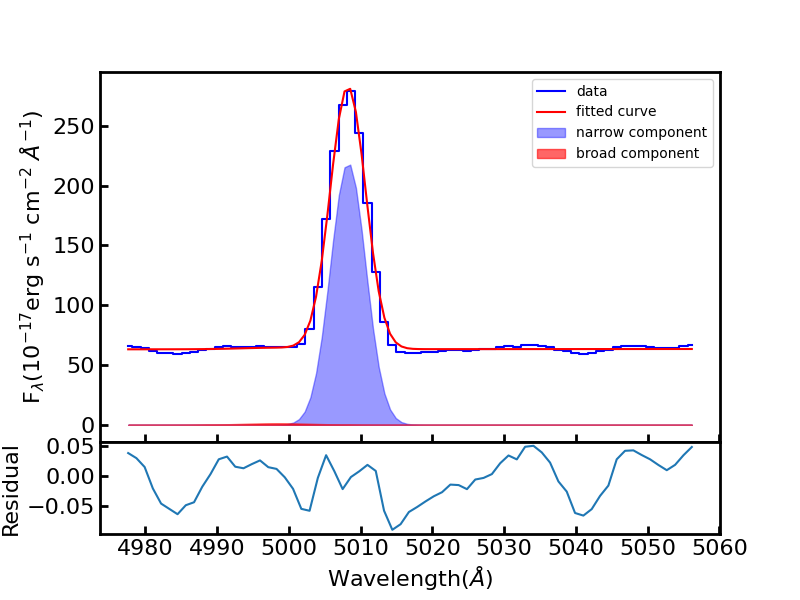}
    \includegraphics[scale=0.3]{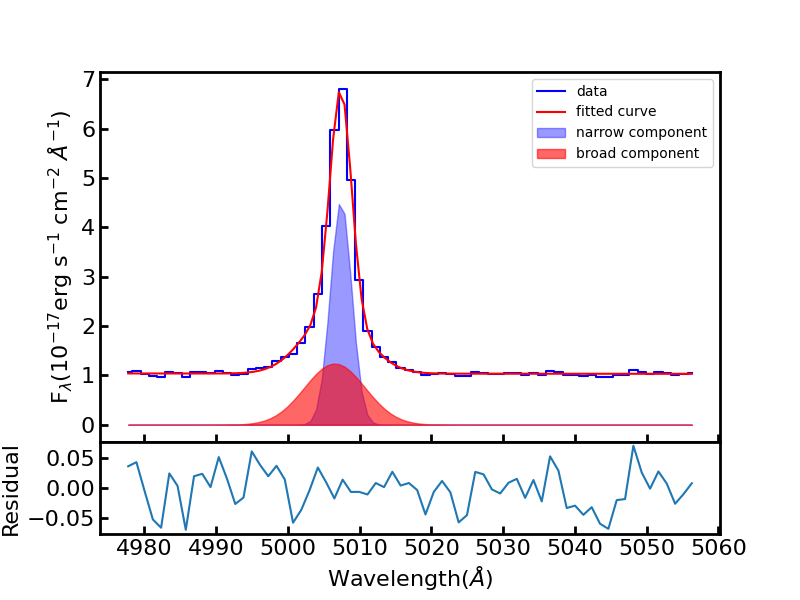}
    \includegraphics[scale=0.3]{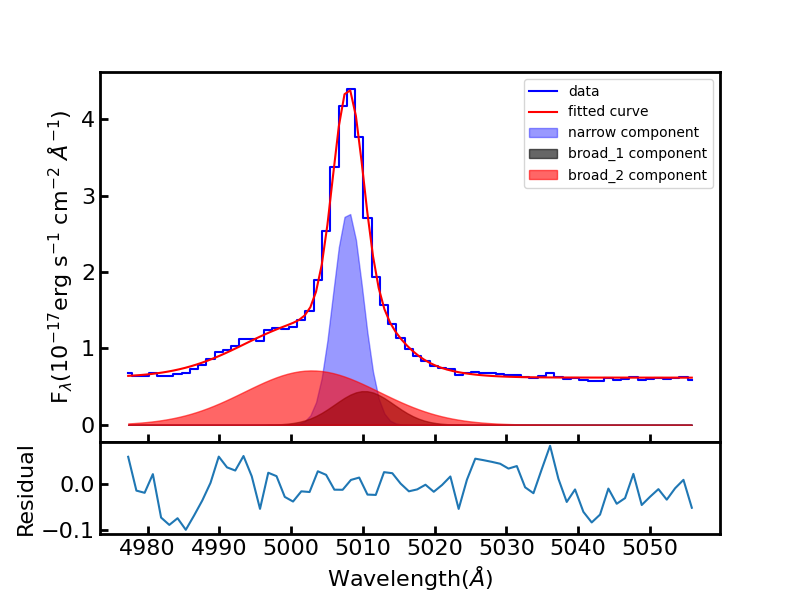}
    }
    \caption{Example line fits to the [OIII]$\lambda$5007 line for an outflow undetected (left panel) and an outflow detected (middle and right panels) source. A single Gaussian profile nicely describes the observed line profile (left panel), while two and three Gaussian components (two phases of outflow) were required for the observed line profiles in the middle and right panels. The broad Gaussian components in the middle and right panels show the presence of outflow. }
    \label{fig:fitting}
\end{figure}

To identify warm ionised outflows in the final
sample of AGN, we carried out Gaussian fits to the [OIII]$\lambda$5007 line. Sample fits are shown in Fig. \ref{fig:fitting}.

\section{Kinematics properties of the less luminous and high-speed outflows}\label{app-table}
To investigate the kinematic properties of the outflows in both the radio-detected and radio-undetected samples (Section \ref{sec:outflow_kine}), we have initially 
used the brightest components, the results for which are given in Table \ref{tab:out_high_lum}. We also carried out a similar analysis using the less-luminous
outflows and the higher velocity outflows, the results of which are given in Tables \ref{tab:out_low_lum} and  \ref{tab:out_high_speed} respectively.
\begin{table}[ht]
    \caption{Kinematic properties of the less luminous outflows of Radio-detected sources (upper panel) and Radio-undetected sources (lower panel)}
    \label{tab:out_low_lum}
    \centering
    \hspace*{-3cm}
    \begin{tabular}{lrrrrrrrrr}
    \hline
    Parameter & \multicolumn{3}{c}{Total} & \multicolumn{3}{c}{Seyferts} & \multicolumn{3}{c}{LINERs}\\
              & Range & Mean & Median & Range & Mean & Median & Range & Mean & Median \\
    \hline
     V$_{shift}$(km s$^{-1}$)  & $-$695.0 to 882.0 &  $-$194.0 &  $-$198.0 & $-$583.0 to 882.0  &  $-$127.0  &  $-$163.0 & $-$695.0 to 258.0  &  $-$376.0  &  $-$449.0 \\
     FWHM$_{out}$(km s$^{-1}$) &  156.0 to 1401.0  &  653.0  & 635.0 & 156.0 to 1401.0  & 644.0  &  619.0 &  257.0 to 1018.0  & 678.0  &  737.0\\
     V$_{out}$(km s$^{-1}$)  & 308.0 to 1503.0 &  823.0  &  818.0 &  308.0 to 1503.0  &  768.0  &  733.0 & 378.0 to 1400.0 & 973.0 &  1084.0 \\
     AI &  $-$0.46 to 0.12 &  $-$0.15 &  $-$0.16 & $-$0.34 to 0.12 &  $-$0.14 & $-$0.15 & $-$0.46 to 0.11 &  $-$0.18 &  $-$0.16 \\
     M$_{out}$(10$^2$ M$_{\odot}$) & 1.11 to 23558.28 & 775.31 & 235.03 &  1.11 to 23558.28 &  960.12  &  348.17 &  1.81 to 2120.98 &  273.7 & 41.16 \\
     $\dot{M}_{out}$(10$^{-3}$ M$_{\odot}$ yr$^{-1}$) &  0.07 to 5112.32  &  145.64  &  33.95 &   0.07 to 5112.32 &  183.5 & 43.07 &  0.33 to 424.39 &  42.87 &  8.02 \\
     KP$_{out}$(10$^{38}$ erg s$^{-1}$) &  0.02 to 18166.25 &  489.61 & 47.73 &  0.02 to 18166.25 &  629.0 & 62.1 &  0.2 to 697.58  &  124.27 &  57.6 \\
     $\dot{P}_{out}$ (10$^{30}$ g cm s$^{-2}$) &  0.14 to 34222.09 &  922.0 &  134.0 &  0.14 to 34222.09 & 1175.24 &  181.62 &  0.91 to 2761.69 & 234.64 &  60.19 \\
    \hline
    \end{tabular}

     \hspace*{-3cm}
    \begin{tabular}{lrrrrrrrrr}
    \hline
    Parameter & \multicolumn{3}{c}{Total} & \multicolumn{3}{c}{Seyferts} & \multicolumn{3}{c}{LINERs}\\
              & Range & Mean & Median & Range & Mean & Median & Range & Mean & Median \\
    \hline
     V$_{shift}$(km s$^{-1}$)   &  $-$657.0 to 227.0 & $-$277.0 &  $-$235.0 & $-$657.0 to 227.0 &  $-$225.0  & -186.0 &  $-$624.0 to $-$211.0 & $-$526.0 &  $-$585.0 \\
     FWHM$_{out}$(km s$^{-1}$)  &  152.0 to 1171.0 &  560.0 &  577.0 & 152.0 to 1151.0 &  527.0 & 505.0 &   580.0 to 1171.0 & 720.0 & 672.0 \\
     V$_{out}$(km s$^{-1}$)     &  237.0 to 1387.0 & 769.0 & 733.0 &  237.0 to 1343.0 & 692.0 &  643.0 & 849.0 to 1387.0 & 1137.0 & 1136.0 \\
     AI   & $-$0.51 to 0.16 & $-$0.18 & $-$0.17 & $-$0.51 to 0.16 & $-$0.15 & $-$0.13 & $-$0.44 to $-$0.01 & $-$0.33 & $-$0.37   \\
     M$_{out}$(10$^2$ M$_{\odot}$) & 0.3 to 1481.81 & 202.86 & 78.37 &   4.7 to 1481.81 & 228.63 & 91.57 & 0.3 to 616.59 & 79.95 &  33.35 \\
     $\dot{M}_{out}$(10$^{-3}$ M$_{\odot}$ yr$^{-1}$) &  0.08 to 247.09 &  24.88 &  9.87 &   1.08 to 247.09 &  26.1 & 11.58 &  0.08 to 151.93 & 19.11 & 7.42 \\
     KP$_{out}$(10$^{38}$ erg s$^{-1}$) &  0.36 to 1053.8 & 59.27 & 14.97 & 0.46 to 1053.8 & 54.11 & 14.42 &  0.36 to 696.0 & 83.9 & 22.41 \\
     $\dot{P}_{out}$ (10$^{30}$ g cm s$^{-2}$) &  0.58 to 1812.07 & 123.71 & 52.05 & 2.63 to 1812.07 &  119.91 & 55.32 & 0.58 1154.74 & 141.85 & 50.95 \\
         \hline
    \end{tabular}

\end{table}

\begin{table}[ht]
    \caption{Kinematic properties of the high-speed outflows of Radio-detected sources (upper panel) and Radio-undetected sources (lower panel)}
    \label{tab:out_high_speed}
    \centering
    \hspace*{-3cm}
    \begin{tabular}{lrrrrrrrrr}
    \hline
    Parameter & \multicolumn{3}{c}{Total} & \multicolumn{3}{c}{Seyferts} & \multicolumn{3}{c}{LINERs}\\
              & Range & Mean & Median & Range & Mean & Median & Range & Mean & Median \\
    \hline
     V$_{shift}$(km s$^{-1}$)  &  $-$782.0 to 882.0 & $-$198.0 & $-$193.0 & $-$782.0 to 882.0 & $-$133.0 & $-$133.0 &
  $-$695.0 to 258.0 & $-$376.0 & $-$449.0 \\
     FWHM$_{out}$(km s$^{-1}$) &  176.0 to 1401.0 & 728.0 & 682.0 &
   176.0 to 1401.0 & 731.0 & 679.0 & 257.0 to 1018.0 & 678.0 & 737.0\\
     V$_{out}$(km s$^{-1}$)  &   337.0 to 1970.0 & 879.0 & 864.0 &  337.0 to 1970.0 & 844.0 & 826.0 & 378.0 to 1400.0 & 973.0 & 1084.0 \\
     AI &  $-$0.46 to 0.12 &  $-$0.15 &  $-$0.16 & $-$0.34 to 0.12 &  $-$0.14 & $-$0.15 & $-$0.46 to 0.11 &  $-$0.18 &  $-$0.16 \\
     M$_{out}$(10$^2$ M$_{\odot}$) &  1.81 to 23558.28 & 831.92 & 239.35 & 1.89 to 23558.28 & 1037.58 & 384.92 & 1.81 to 2120.98 & 273.7 & 41.16 \\
     $\dot{M}_{out}$(10$^{-3}$ M$_{\odot}$ yr$^{-1}$) &  0.23 to 5112.32 & 163.81 &  42.14 &  0.23 to 5112.32 & 208.36 & 56.62 & 0.33 to 424.39 & 42.87 & 8.02 \\
     KP$_{out}$(10$^{38}$ erg s$^{-1}$) &  0.2 to 18166.25 & 602.86 & 62.1 & 0.24 to 18166.25 & 783.98 & 101.5 & 0.2 to 1425.12 & 111.26  &  34.4  \\
     $\dot{P}_{out}$ (10$^{30}$ g cm s$^{-2}$) & 0.83 to 34222.09 & 1079.06 & 180.65 & 0.83 to 34222.09 & 1390.16 &  267.04 & 0.91 to 2761.69 & 234.64 & 60.19 \\
    \hline
    \end{tabular}

     \hspace*{-3cm}
    \begin{tabular}{lrrrrrrrrr}
    \hline
    Parameter & \multicolumn{3}{c}{Total} & \multicolumn{3}{c}{Seyferts} & \multicolumn{3}{c}{LINERs}\\
              & Range & Mean & Median & Range & Mean & Median & Range & Mean & Median \\
    \hline
     V$_{shift}$(km s$^{-1}$)   & $-$657.0 to 177.0 & $-$278.0 & $-$235.0 &
  $-$657.0 to 177.0 & $-$225.0 & $-$185.0 & $-$624.0 to $-$211.0  &  $-$526.0 & $-$585.0 \\
     FWHM$_{out}$(km s$^{-1}$)  &  172.0 to 1171.0 &  575.0 &  580.0 &
 172.0 to 1151.0 & 545.0 & 536.0 &  580.0 to 1171.0 & 720.0 &  672.0  \\
     V$_{out}$(km s$^{-1}$)     & 312.0 to 1387.0 & 777.0 & 733.0 & 312.0 to 1343.0 & 701.0 & 657.0 & 849.0 to 1387.0 & 1137.0 & 1136.0  \\
     AI   & $-$0.51 to 0.16 & $-$0.18 & $-$0.17 & $-$0.51 to 0.16 & $-$0.15 & $-$0.13 & $-$0.44 to $-$0.01 & $-$0.33 & $-$0.37   \\
     M$_{out}$(10$^2$ M$_{\odot}$) &  0.3 to 1488.25 & 207.14 &  84.41 &  4.7 to 1488.25 & 233.8 & 94.93 & 0.3 to 616.59 & 79.95 & 33.35 \\
     $\dot{M}_{out}$(10$^{-3}$ M$_{\odot}$ yr$^{-1}$) &  0.08 to 247.09 & 26.72 & 9.97 & 1.08 to 247.09 & 28.32 & 11.58 &
   0.08 to 151.93 &  19.11 & 7.42  \\
     KP$_{out}$(10$^{38}$ erg s$^{-1}$) & 0.36 to 1053.8 & 63.07 & 14.99 &  0.46 to 1053.8 & 58.71 & 14.98 & 0.36 to 696.0 & 83.9 & 22.41 \\
     $\dot{P}_{out}$ (10$^{30}$ g cm s$^{-2}$) & 0.58 to 1812.07 & 133.73 & 52.05 &  2.63 to 1812.07 &  132.02 &  55.32 &
 0.58 to 1154.74 & 141.85 & 50.95 \\
         \hline
    \end{tabular}

\end{table}

\clearpage
\bibliography{ref}{}
\bibliographystyle{aasjournal}



\end{document}